\documentclass[preprint,showpacs,preprintnumbers,amsmath,amssymb]{revtex4}
\pdfoutput=1

\usepackage{graphicx}
\usepackage{dcolumn}
\usepackage{bm}
\usepackage{amssymb}

\begin{document}
\title{Classical double-well systems coupled to finite baths
}

\author{Hideo Hasegawa}
\altaffiliation{hideohasegawa@goo.jp}
\affiliation{Department of Physics, Tokyo Gakugei University,  
Koganei, Tokyo 184-8501, Japan}%

\date{\today}

\begin{abstract}
We have studied properties of a classical $N_S$-body double-well system
coupled to an $N_B$-body bath, performing simulations of $2(N_S+N_B)$ first-order 
differential equations with $N_S \simeq 1 - 10$ and $N_B \simeq 1 - 1000$.
A motion of Brownian particles in the absence of external forces 
becomes chaotic for appropriate model parameters such as $N_B$, $c_o$ (coupling strength), 
and $\{ \omega_n\}$ (oscillator frequency of bath): 
For example, it is chaotic for a small $N_B$ ($\lesssim 100$) but regular for a large $N_B$ ($\gtrsim 500$).
Detailed calculations of the stationary energy distribution of the system $f_S(u)$
($u$: an energy per particle in the system)
have shown that its properties are mainly determined
by $N_S$, $c_o$ and $T$ (temperature) but weakly depend on $N_B$ and $\{\omega_n \}$.
The calculated $f_S(u)$ is analyzed with the use of the $\Gamma$ distribution. 
Difference and similarity between properties of double-well and harmonic-oscillator systems
coupled to finite bath are discussed.

\end{abstract}

\pacs{05.40.-a, 05.70.-a, 05.10.Gg}
        

\maketitle
\newpage
\section{Introduction}

Many studies have been made with the use of a model describing a classical 
or quantum open system which is coupled to baths consisting of a collection of
harmonic oscillators. Such a model is conventionally referred 
to as the Caldeira-Leggett (CL) model 
\cite{Caldeira81,Caldeira83},  
although equivalent models had been proposed earlier 
by Magalinskii \cite{Magalinskii59} and Ullersma \cite{Ullersma66}.
From the CL model, we may derive 
the Langevin model with dissipation and diffusion (noise) terms.
Originally the CL model was introduced for $N_B$-body bath with  
$N_B \rightarrow \infty$, for which the Ohmic and Drude-type spectral densities with
continuous distributions are adopted. 
Furthermore in the original CL model, the number of particles in a systems, $N_S$, 
is taken to be unity ($N_S=1$). We expect that
a generic open system may contain any number of particles and that a system may 
be coupled to a bath consisting of finite harmonic oscillators in general.
In recent years, the CL model has been employed for a study of properties 
of open systems with finite $N_S$ and/or $N_B$  
\cite{Smith08,Wei09,Hanggi08,Ingold09,Rosa08,Hasegawa11,Carcaterra11,Hasegawa11c,Hasegawa11b}.
Specific heat anomalies of quantum oscillator (system) coupled to finite bath
have been studied \cite{Hanggi08,Ingold09,Hasegawa11c}.
A thermalization \cite{Smith08,Wei09}, energy exchange \cite{Rosa08},
dissipation \cite{Carcaterra11} and the Jarzynski equality \cite{Jarzynski97,Hasegawa11b}
in classical systems coupled to finite bath have been investigated.

In a previous paper \cite{Hasegawa11}, we have studied the $(N_S+N_B)$ model
for finite $N_S$-body systems coupled to baths consisting of $N_B$
harmonic oscillators. Our study for open harmonic oscillator systems 
with $N_S \simeq 1-10$ and $N_B \simeq 10-1000$ has shown that stationary energy
distribution of the system has a significant and peculiar dependence on $N_S$, 
but it weakly depends on $N_B$ \cite{Hasegawa11}.
These studies mentioned above 
\cite{Smith08,Wei09,Hanggi08,Ingold09,Rosa08,Hasegawa11,Carcaterra11,Hasegawa11c,Hasegawa11b}
have been made for harmonic-oscillator systems with finite $N_S$ and/or $N_B$.   

Double-well potential models have been employed in a wide range of
fields including physics, chemistry and biology 
(for a recent review on double-well system, see Ref. \cite{Thorwart01}).
Various phenomena such as the stochastic resonance (SR), tunneling through potential barrier
and thermodynamical properties \cite{Hasegawa12b} have been studied. 
The CL model for the double-well systems with $N_S=1$ and
$N_B = \infty$ has been extensively employed for a study on the SR \cite{Gamma98}. 
Properties of SR for variations of magnitude of white noise 
\cite{Hanggi84,Hanggi93,Gamma89,Gamma98}
and relaxation time of colored noise \cite{Neiman96,Hasegawa12} have been studied.
However, studies for open double-well systems with finite $N_S$ and/or $N_B$
have not been reported as far as we are aware of.
It would be interesting and worthwhile to study open classical double-well systems
described by the $(N_S+N_B)$ model with finite $N_S$ and $N_B$, 
which is the purpose of the present paper.

The paper is organized as follows.
In Sec. II, we briefly explain the $(N_S+N_B)$ model proposed in our previous
study \cite{Hasegawa11}.
In Sec. III, direct simulations (DSs) of $2(N_S+N_B)$ first-order differential equations
for the adopted model have been performed.
Dynamics of a single double-well system ($N_S=1$) 
coupled to a finite bath ($2 \leq N_B \leq 1000$) in the phase space 
is investigated (Sec. III B). 
We study stationary energy distributions in the system and bath, performing
detailed DS calculations, changing $N_S$, $N_B$, the coupling strength and
the distribution of bath oscillators (Sec. III C).
Stationary energy and position distributions obtained by DSs are analyzed in Sec. IV.
The final Sec. V is devoted to our conclusion.

\section{Adopted ($N_S+N_B$) model}
We consider a system including $N_S$ Brownian particles coupled to
a bath consisting of independent $N_B$ harmonic oscillators.
We assume that the total Hamiltonian is given by \cite{Hasegawa11}
\begin{eqnarray}
H &=& H_S+H_B+H_{I},
\label{eq:A0}
\end{eqnarray}
with
\begin{eqnarray}
H_S &=& \sum_{k=1}^{N_S} \left[ \frac{P_k^2}{2M}+ V(Q_k) \right],
\label{eq:A2}\\
H_B &=& \sum_{n=1}^{N_B} \left[ \frac{p_n^2}{2m}+ \frac{m \omega_n^2}{2} q_n^2 \right], 
\label{eq:A3}\\
H_{I} &=& \frac{1}{2} \sum_{k=1}^{N_S}  \sum_{n=1}^{N_B} 
c_{kn} (Q_k-q_n)^2, 
\label{eq:A1}
\end{eqnarray}
where $H_S$, $H_B$ and $H_I$ express Hamiltonians for the system, bath 
and interaction, respectively.
Here $M$ ($m$) denotes the mass, $P_k$ ($p_n$) the momentum, 
$Q_k$ ($q_n$) position of the oscillator in the system (bath),
$V(Q_k)$ signifies the potential in the system, $\omega_n$ stands for 
oscillator frequency in the bath, and $c_{nk}$ is coupling constant. 
The model is symmetric with respect to an exchange of system $ \leftrightarrow $ bath
if $V(Q)$ is the harmonic potential.
From Eqs. (\ref{eq:A0})-(\ref{eq:A1}), we obtain $2(N_S+N_B)$ first-order
differential equations, 
\begin{eqnarray}
\dot{Q}_k &=& \frac{P_k}{M}, 
\label{eq:A10b}\\
\dot{P}_k 
&=& - V'(Q_k)-\sum_{n=1}^{N_B} c_{kn} (Q_k-q_n), 
\label{eq:A10c}\\
\dot{q}_n &=& \frac{p_n}{m}, \\
\dot{p}_n
&=& -m \omega_n^2 q_n- \sum_{k=1}^{N_S} c_{kn} (q_n-Q_k),
\label{eq:A11b}
\end{eqnarray}
which yield
\begin{eqnarray}
M \ddot{Q}_k 
&=& - V'(Q_k)-\sum_{n=1}^{N_B} c_{kn} (Q_k-q_n), 
\label{eq:A10}\\
m \ddot{q}_n
&=& -m \omega_n^2 q_n- \sum_{k=1}^{N_S} c_{kn} (q_n-Q_k),
\label{eq:A11}
\end{eqnarray}
with prime ($'$) and dot ($\cdot$) denoting derivatives with respect 
to the argument and time, respectively.
It is noted that the second term of Eq. (\ref{eq:A10c}) or (\ref{eq:A10}) given by
\begin{eqnarray}
F_k^{(eff)} &=& -\sum_{n=1}^{N_B} c_{kn} (Q_k-q_n),
\label{eq:E4}
\end{eqnarray}
plays a role of the effective force to the $k$th system.

A formal solution of Eq. (\ref{eq:A11}) for $q_n(t)$ is given by
\begin{eqnarray}
q_n(t) &=& q_n(0) \cos \tilde{\omega}_n t 
+ \frac{\dot{q}_n(0)}{\tilde{\omega}_n} \sin \tilde{\omega}_n t 
+ \sum_{\ell=1}^{N_S} \frac{c_{\ell n}}{m \tilde{\omega}_n} \int_0^t 
\sin \tilde{\omega}_n (t-t') Q_{\ell}(t')\:dt',
\label{eq:A12}
\end{eqnarray}
with
\begin{eqnarray}
\tilde{\omega}_n^2 &=&  \frac{b_n}{m}+\sum_{k=1}^{N_S} \frac{c_{kn}}{m}
=\omega_n^2 +\sum_{k=1}^{N_S} \frac{c_{kn}}{m}.
\end{eqnarray}
Substituting Eq. (\ref{eq:A12}) to Eq. (\ref{eq:A10}), 
we obtain the non-Markovian Langevin equation given by 
\begin{eqnarray}
M \ddot{Q}_k(t) &=& -V'(Q_k)- M \sum_{\ell=1}^{N_S} \xi_{k\ell} Q_{\ell}(t)
- \sum_{\ell=1}^{N_S} \int_0^t \gamma_{k\ell}(t-t') \dot{Q}_{\ell}(t')\:dt' 
\nonumber \\
&-& \sum_{\ell=1}^{N_S} \gamma_{k\ell}(t) Q_{\ell}(0) + \zeta_k(t)
\hspace{1cm}\mbox{($k =1$ to $N_S$)},
\label{eq:A13}
\end{eqnarray} 
with
\begin{eqnarray}
M \xi_{k\ell} &=& \sum_{n=1}^{N_B}  \left[ c_{kn}  \delta_{k \ell} 
-\frac{c_{kn} c_{\ell n}}{m \tilde{\omega}_n^2} \right], 
\label{eq:A14}\\
\gamma_{k\ell}(t) 
&=&\sum_{n=1}^{N_B} \left( \frac{c_{kn} c_{\ell n}}{m \tilde{\omega}_n^2}\right) 
\cos \tilde{\omega}_n t, 
\label{eq:A15}\\
\zeta_k(t) &=& \sum_{n=1}^{N_B} c_{kn} 
\left[q_n(0) \cos \tilde{\omega}_n t
+ \frac{\dot{q}_n(o)}{\tilde{\omega}_n} \sin \tilde{\omega}_n t \right],
\label{eq:A16}
\end{eqnarray}
where $\xi_{k \ell}$ denotes the additional interaction between
$k$ and $\ell$th particles in the system induced by couplings $\{ c_{kn} \}$, 
$\gamma_{k\ell}(t)$ the memory kernel and $\zeta_k$ the stochastic force.

If the equipartition relation is realized 
in initial values of $q_n(0)$ and $\dot{q}(0)$,
\begin{eqnarray}
\langle m \tilde{\omega}_n^2 q_n(0)^2\rangle_B
&=& \langle m  \dot{q}_n(0)^2\rangle_B = k_B T,
\label{eq:A17}
\end{eqnarray} 
we obtain the fluctuation-dissipation relation:
\begin{eqnarray}
\langle \zeta_k(t) \zeta_k(t') \rangle_B &=& k_B T \gamma_{kk}(t-t'),
\label{eq:A18}
\end{eqnarray}
where $\langle \cdot \rangle_B$ stands for the average
over variables in the bath.

In the case of $N_B \rightarrow \infty$, summations 
in Eqs. (\ref{eq:A14})-(\ref{eq:A16}) are replaced by integrals. 
When the spectral density defined by
\begin{eqnarray}
J(\omega)= \frac{\pi}{2} \sum_n \frac{c_n^2}{m_n \omega_n^2} \delta(\omega-\omega_n),
\label{eq:A21}
\end{eqnarray}
is given by the Ohmic form: $J(\omega) \propto \omega$ for $0 \leq \omega < w_D$, 
the kernel becomes
\begin{eqnarray}
\gamma(t) \propto \frac{\sin \omega_D t}{\pi t} \propto \delta(t),
\end{eqnarray}
which leads to the Markovian Langevin equation.

In the case of $N_S=1$, we obtain $\xi$ and $\gamma$ 
in Eqs. (\ref{eq:A14}) and (\ref{eq:A15}) 
where the subscripts $k$ and $\ell$ are dropped ({\it e.g.,} $c_{kn}=c_n$),
\begin{eqnarray}
M \xi(t) &=& \sum_{n=1}^{N_B} c_n \left( 1-\frac{c_n}{m \tilde{\omega}_n^2} \right), 
\label{eq:A19}\\
\gamma(t) &=& \sum_{n=1}^{N_B} \left( \frac{c_n^2}{m \tilde{\omega}_n^2}\right) \cos \tilde{\omega}_n t.
\label{eq:A20}
\end{eqnarray}
The additional interaction vanishes ($\xi=0$)
if we choose $c_n=m \tilde{\omega}_n^2$ in Eq. (\ref{eq:A19}). 

In the case of $N_S \neq 1$, however, it is impossible to choose
$\{ c_{kn} \}$ such that $\xi_{k \ell}=0$ is realized for all pairs of $(k, \ell)$
in Eq. (\ref{eq:A14}).
Then $Q_k$ is inevitably coupled to $Q_{\ell}$ for $\ell \neq k$ 
with the superexchange-type interaction of antiferromagnets: 
$-\sum_n c_{kn}c_{\ell n}/m \tilde{\omega}_n^2$ in Eq. (\ref{eq:A14}).

\section{Model calculations for double-well systems}
\subsection{Calculation methods}

We consider a system with the double-well potential
\begin{eqnarray}
V(Q) &=& \left( \frac{\Delta}{Q_0^4} \right) (Q^2-Q_0^2)^2,
\label{eq:E3}
\end{eqnarray}
which has the stable minima 
of $V(\pm Q_0)=0$ at $Q=\pm Q_0$ and locally unstable maximum
of $V(0)=\Delta$ at $Q=0$ with the barrier height $\Delta$.
We have adopted $Q_0=1.0$ and $\Delta=1.0$ in our DSs.

It is easier to solve $2(N_S+N_B)$ first-order differential equations
given by Eqs. (\ref{eq:A10b})-(\ref{eq:A11b}) 
than to solve the $N_S$ Langevin equations given by Eqs. (\ref{eq:A13})-(\ref{eq:A16})
although the latter provides us with clearer physical insight than the former. 
In order to study the $N_S$ and $N_B$ dependences of various physical quantities,
we have assumed that the coupling $c_{k n}$ is given by \cite{Hasegawa11,Note2}
\begin{eqnarray}
c_{k n}=\frac{c_o}{N_S N_B},
\label{eq:E1}
\end{eqnarray}
because the interaction term includes summations of 
$\sum_{k=1}^{N_S}$ and $\sum_{n=1}^{N_B}$ in Eq. (\ref{eq:A1}).
It is noted that with our choice of $c_{k n}$, the interaction contribution
is finite even in the thermodynamical limit of $N_B \rightarrow \infty$ because
the summation over $n$ runs from 1 to $N_B$ in Eq. (\ref{eq:A1}).
DSs of Eqs. (\ref{eq:A10b})-(\ref{eq:A11b}) have been performed with the use 
of the fourth-order Runge-Kutta method with the time step of 0.01.
We have adopted $k_B=1.0$, $M=m=1.0$, $c_o=1.0$, and $\omega_n=1.0$ otherwise noticed.

We consider energies per particle $u_{\eta}(t)$ 
in the system ($\eta$=S) and the bath ($\eta$=B) which are assume to be given by
\begin{eqnarray}
u_S &=& \frac{1}{N_S} \sum_{k=1}^{N_S} \left[\frac{P_k^2}{2 M}
+ V(Q_k) \right], \\
u_B &=& \frac{1}{N_B} \sum_{n=1}^{N_B} \left[\frac{p_n^2}{2 m}
+ \frac{m \omega_n^2 q_n^2}{2} \right],
\end{eqnarray}
which is valid for the weak interaction, although
a treatment of the finite interaction is ambiguous and controversial
\cite{Hanggi08,Ingold09}.

\subsection{Dynamics of a particle in the $(Q, P)$ phase space}

\subsubsection{Effect of $c_o$}
\begin{figure}
\begin{center}
\includegraphics[keepaspectratio=true,width=100mm]{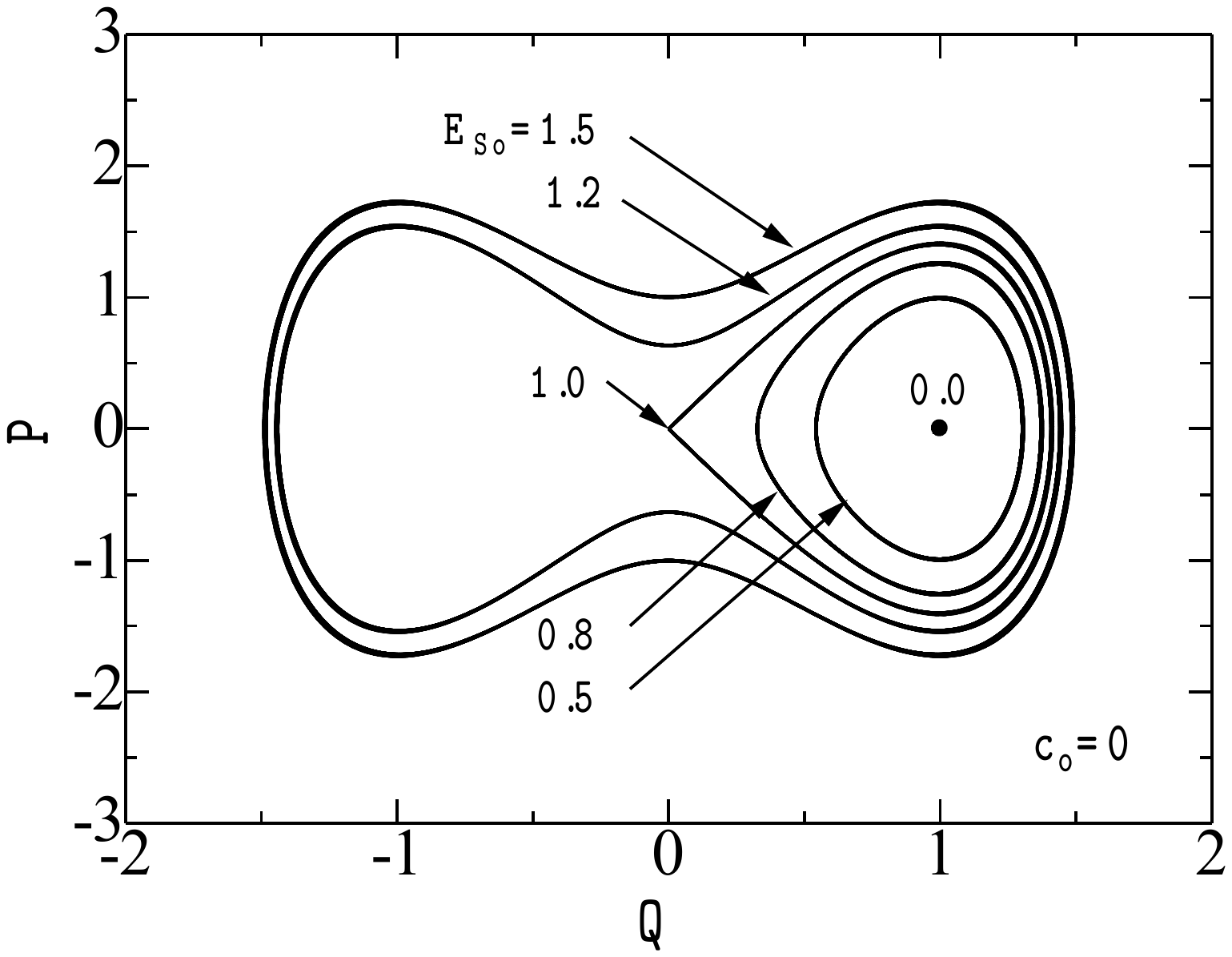}
\end{center}
\caption{
Plot of phase-space trajectories for a particle
in an isolated double-well system ($c_o=0.0$). Trajectories are plotted
for energies of $E_{So}/\Delta=0.0$, 0.5, 0.8, 1.0, 1.2 and 1.5. 
}
\label{fig1}
\end{figure}

First we consider an isolated double-well system ($N_S=1$ and $c_o=0.0$).
Figure \ref{fig1} shows the phase-space trajectories in the $(Q, P)$ phase space for this system with
six different initial system energies $E_{So}$. 
For $E_{So}= 0.0$, the system has two stable fixed points 
at $(Q, P)=(\pm 1.0, 0.0)$, and for $E_{So} = \Delta=1.0$ it has one unstable fixed point
at $(Q, P)=(0.0, 0.0)$. In the case of $0.0 < E_{So} < 1.0$, the trajectory
is restricted in the region of $Q > 0.0$ (or $Q < 0.0$).  
In contrast in the case of $E_{So} > 1.0$,
trajectory may visit both regions of $Q > 0.0$ and $Q < 0.0$.
The case of $E_{So}=1.0$ is critical between the two cases.

\begin{figure}
\begin{center}
\includegraphics[keepaspectratio=true,width=120mm]{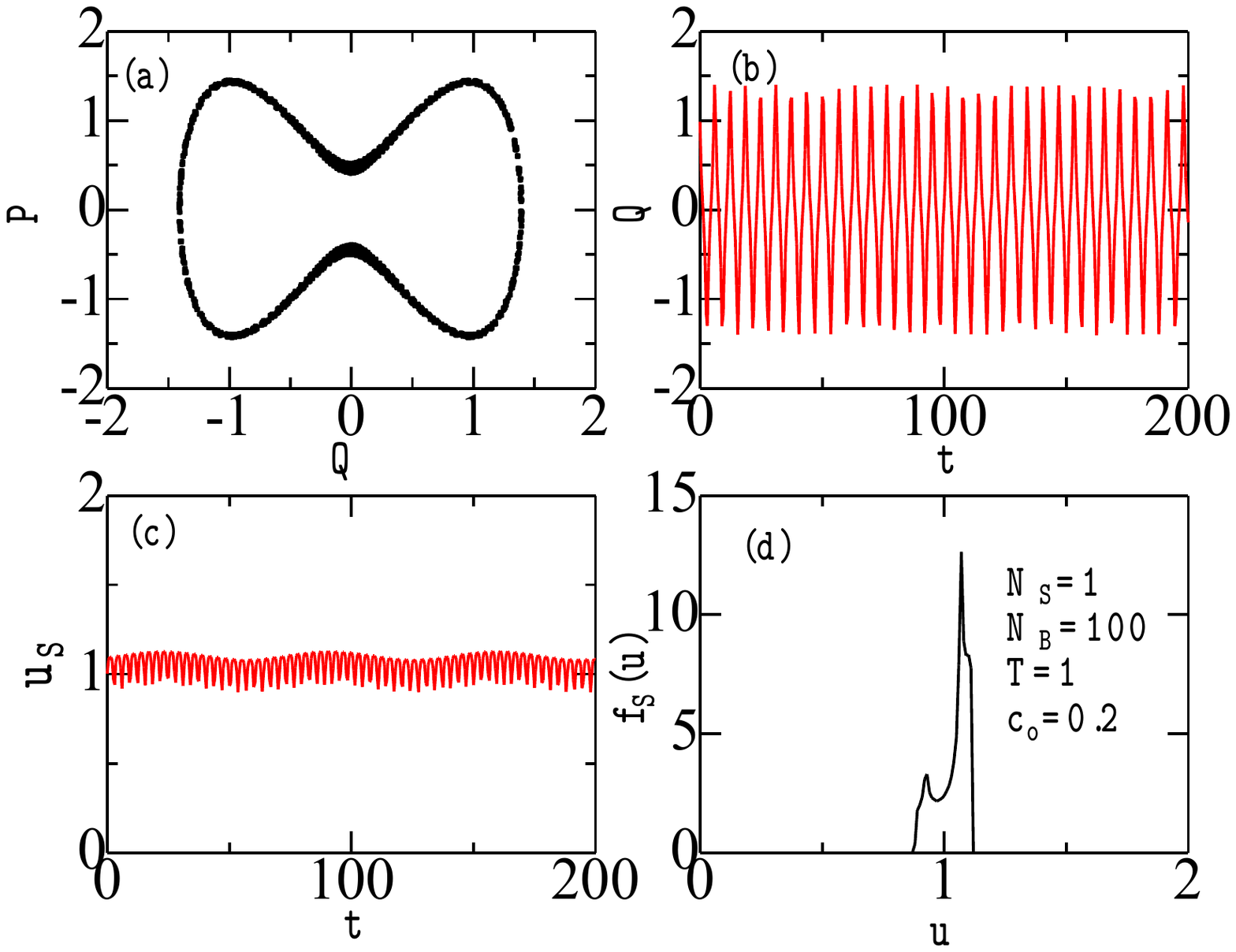}
\end{center}
\caption{
(Color online) 
(a) Strobe plot in the $(Q, P)$ phase space (with a time interval of 1.0),
(b) $Q(t)$, (c) $u_S(t)$, and (d) the system energy distribution $f_S(u)$
obtained by a single run for $E_{So}=1.0$, $N_S=1$, $N_B=100$, $T=1.0$ and $c_o=0.2$. 
}
\label{fig2}
\end{figure}

\begin{figure}
\begin{center}
\includegraphics[keepaspectratio=true,width=120mm]{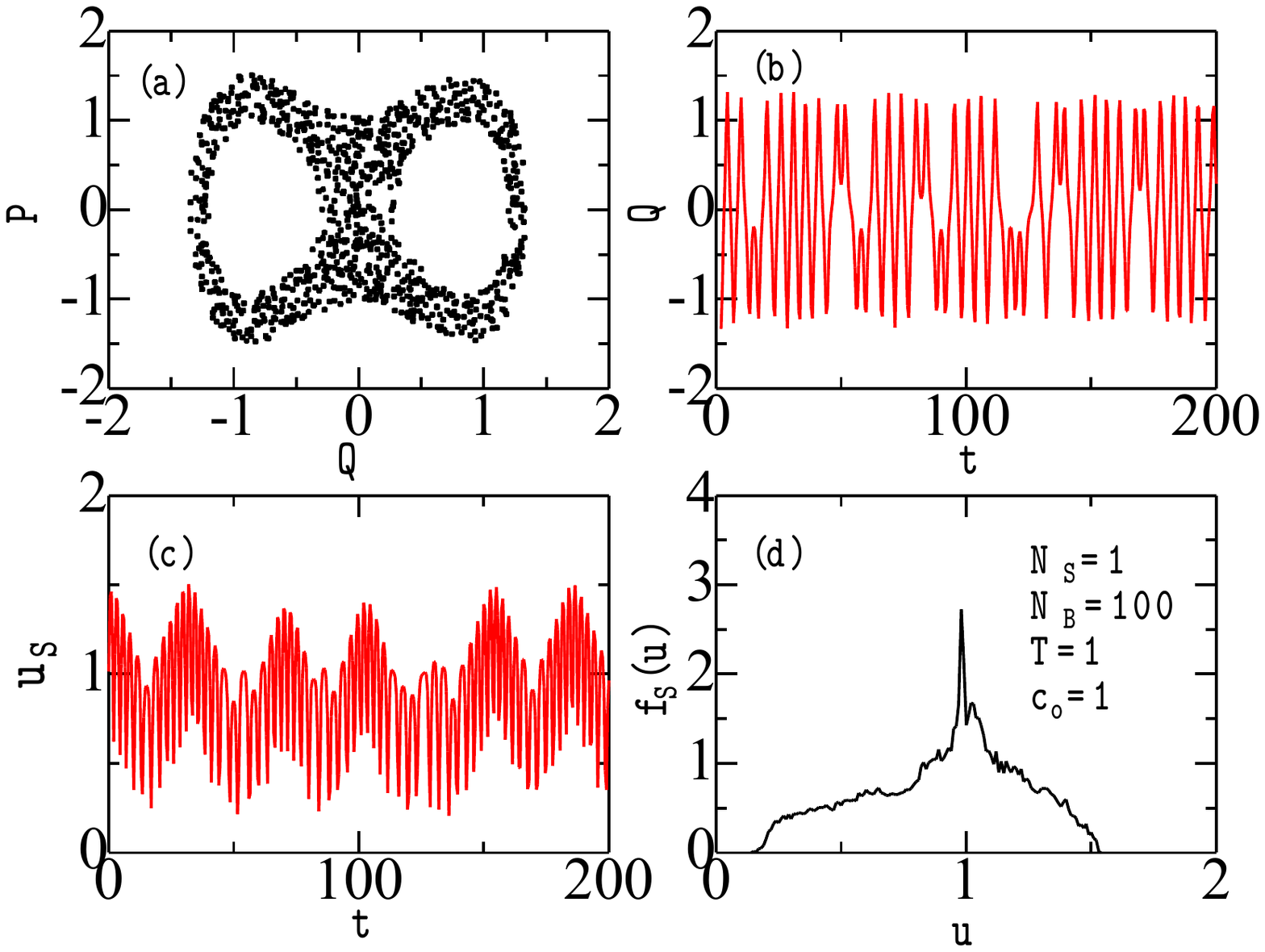}
\end{center}
\caption{
(Color online) 
(a) Strobe plot in the $(Q, P)$ phase space, (b) $Q(t)$, (c) $u_S(t)$, 
and (d) the system energy distribution $f_S(u)$ obtained by a single run
for $E_{So}=1.0$, $N_S=1$, $N_B=100$, $T=1.0$ and $c_o=1.0$. 
}
\label{fig3}
\end{figure}

Next the double-well system is coupled to a bath. In our DSs, 
we  have assumed that system and bath are decoupled at $t < 0$
where they are in equilibrium states with $E_{So}=T$, the temperature $T$ being defined by $T=u_B$.
We have chosen initial values of $Q(0)=1.0$ and $P(0)=\sqrt{2 M [E_{So}-V(Q(0))]}$ 
for a given initial system energy $E_{So}$.
Initial conditions for $q_n(0)$ and $p_n(0)$ are given by random 
Gaussian variables with zero means and variance proportional to $T$ 
[Eq. (\ref{eq:A17})] \cite{Hasegawa11}.
Results to be reported in this subsection have been obtained by single runs for
$t=0$ to 1000.

Figures \ref{fig2}(a) and \ref{fig2}(b) show a strobe plot in the $(Q, P)$ phase space 
(with a time interval of 1.0) and the time-dependence of $Q(t)$, respectively, 
for $E_{So}=1.0$, $N_S=1$, $N_B=100$ and $c_o=0.2$. 
The trajectory starting from $Q(0)=1.0$ goes to the negative-$Q$ region 
because a particle may go over the potential barrier with a help
of a force (noise) originating from bath given by Eq. (\ref{eq:E4}).
The system energy fluctuates as shown in Fig. \ref{fig2}(c), whose distribution 
is plotted in Fig. \ref{fig2}(d). 

Results in Fig. \ref{fig2} are regular. In contrast, when a coupling strength
is increased to $c_o=1.0$, the system becomes chaotic as shown in
Figs. \ref{fig3}(a) and \ref{fig3}(b) where a strobe plot in the $(Q, P)$ phase space and 
the time-dependence of $Q(t)$ are plotted, respectively. 
This is essentially the force-induced chaos in classical double-well system \cite{Reichl84}: 
although an external force is not applied to our system, 
a force arising from a coupling with bath given by Eq. (\ref{eq:E4}) 
plays a role of an effective external force for the system.
Figures \ref{fig3}(c) and \ref{fig3}(d) show that in the case of $c_o=1.0$, 
$u_S$ has more appreciable temporal fluctuations with a wider energy distribution 
in $f_S(u)$ than in the case of $c_o=0.2$.
Although system energies fluctuate, they are not dissipative at 
$0.0 \leq t < 1000.0$ in DSs both for $c_o=0.2$ and $c_o=1.0$ with $N_B=100$. 

\subsubsection{Effect of $\omega_n$ distributions}

\begin{figure}
\begin{center}
\includegraphics[keepaspectratio=true,width=120mm]{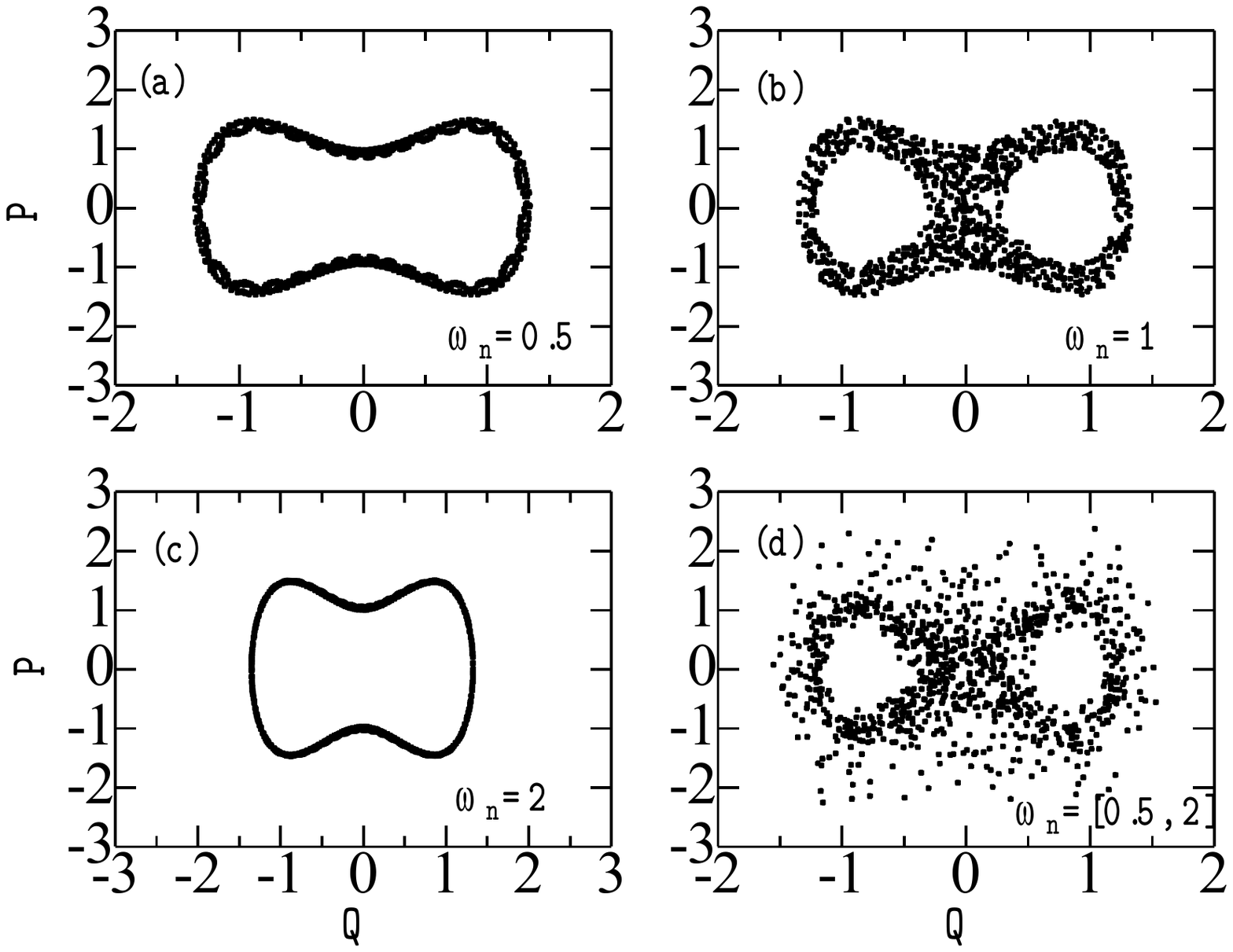}
\end{center}
\caption{
Strobe plots in the $(Q, P)$ phase space for various distribution of $\{ \omega_n \}$:
(a) $\omega_n=0.5$, (b) $\omega_n=1.0$, (c) $\omega_n=2.0$ and (d) $\omega_n \in [0.5, 2.0]$ 
obtained by single runs with $E_{So}=1.0$, $N_S=1$, $N_B=100$, $T=1.0$ and $c_o=1.0$. 
}
\label{fig4}
\end{figure}

We have so far assumed $\omega_n=1.0$ in the bath, which is now changed.
Figures \ref{fig4}(a) and \ref{fig4}(c) show strobe plots for $\omega_n=0.5$
and $2.0$, respectively, which are regular and which are different from 
a chaotic result for $\omega_n=1.0$ shown in Fig. \ref{fig4}(b).
When we adopt $\{ \omega_n \}$ which is randomly distributed in $[0.5, 2.0]$,
a motion of a system particle becomes chaotic as shown in Fig. \ref{fig4}(d).
This is because contributions from $\omega_n \sim 1.0$ among $[0.5. 2.0]$
induce chaotic behavior.

\subsubsection{Effect of $N_B$}

\begin{figure}
\begin{center}
\includegraphics[keepaspectratio=true,width=120mm]{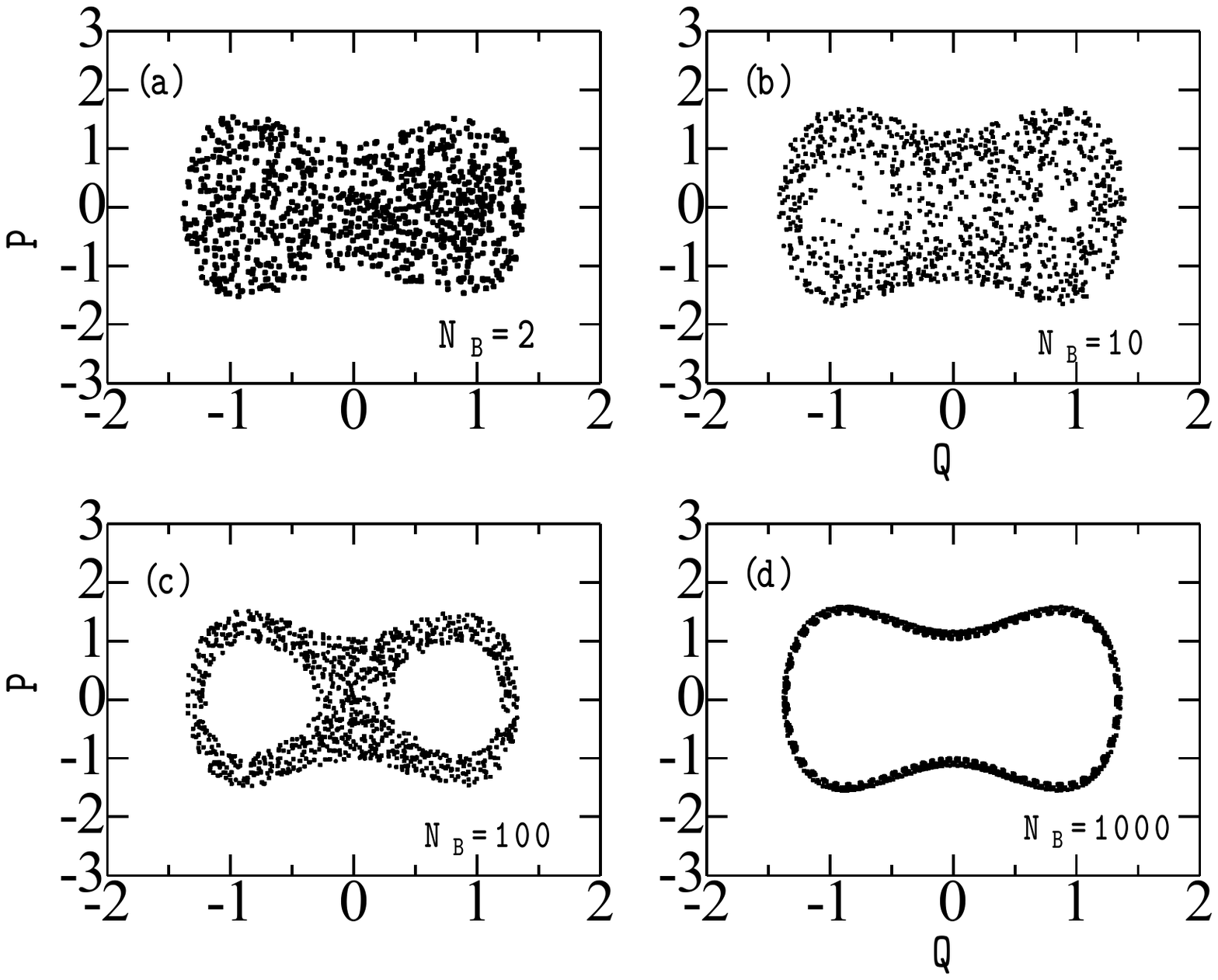}
\end{center}
\caption{
Strobe plots in the $(Q, P)$ phase space for various $N_B$: (a) $N_B=2$, (b) $N_B=10$, 
(c) $N_B=100$ and (d) $N_B=1000$ with $E_{So}=1.0$, $N_S=1$, $T=1.0$ and $c_o=1.0$. 
}
\label{fig5}
\end{figure}

We have repeated calculations by changing $N_B$, whose results are
plotted in Figs. \ref{fig5}(a)-\ref{fig5}(d).
Figures \ref{fig5}(a), \ref{fig5}(b) and \ref{fig5}(c) show that 
chaotic behaviors for $N_B=2$ and $N_B=10$ are more significant than that for $N_B=100$.
On the contrary, chaotic behavior is not realized for $N_B=1000$ in Fig. \ref{fig5}(d),
which is consistent with the fact that chaos has not been reported
for the double-well system subjected to infinite bath.

\subsubsection{Effect of initial system energy $E_{So}$}

\begin{figure}
\begin{center}
\includegraphics[keepaspectratio=true,width=120mm]{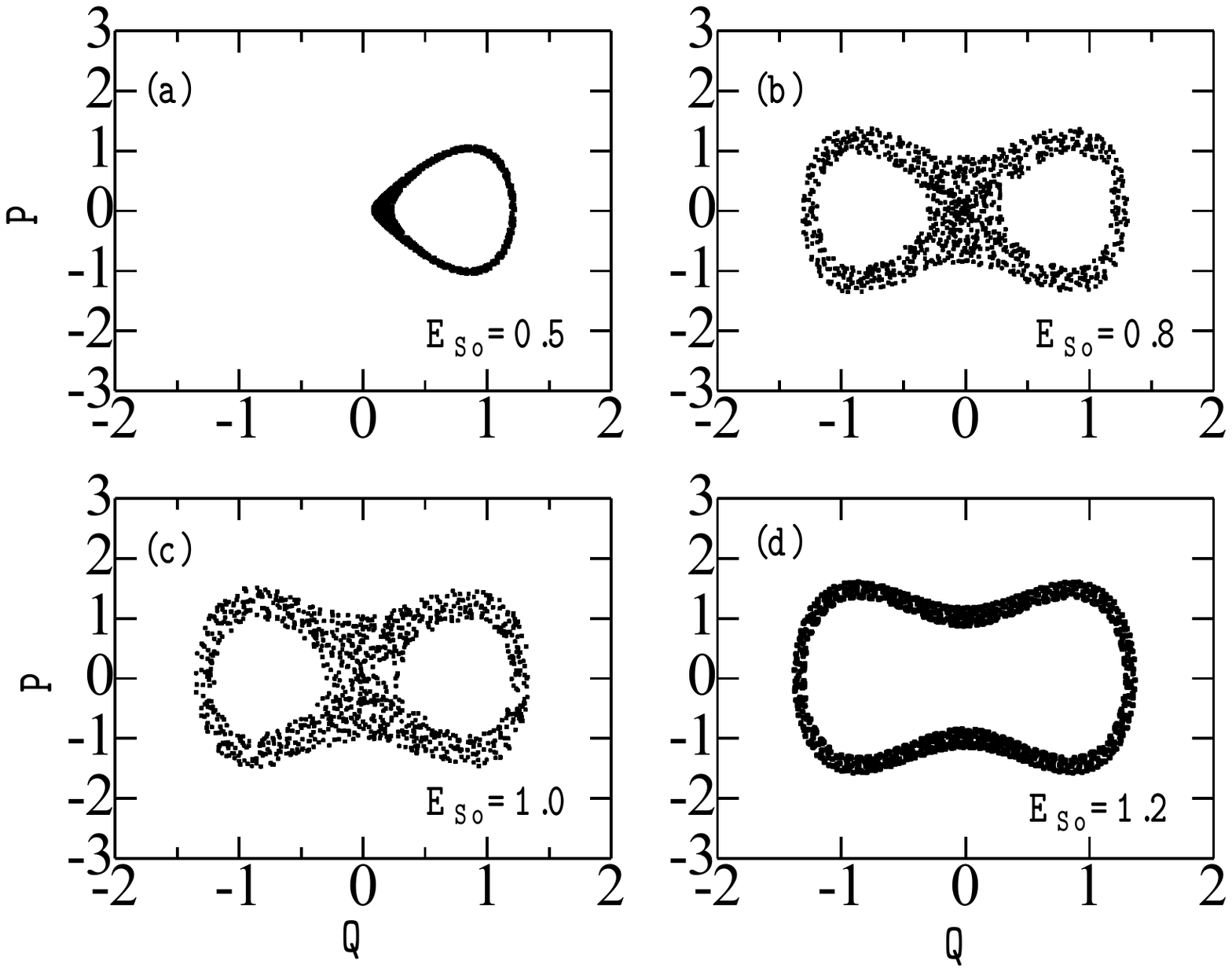}
\end{center}
\caption{
Strobe plots in the $(Q, P)$ phase space for various $E_{So}$: (a) $E_{So}=0.5$, (b) $E_{So}=0.8$ 
(c) $E_{So}=1.0$ and $E_{So}=1.2$ with $N_S=1$, $N_B=100$, $T=1.0$ and $c_o=1.0$. 
}
\label{fig6}
\end{figure}

Next we change the initial system energy of $E_{So}$.
Figures \ref{fig6}(a)-(d) show strobe plots in the $(Q, P)$ phase space
for various $E_{So}$ with $N_S=1$, $N_B=100$, $T=1.0$ and $c_o=1.0$.
Figure  \ref{fig6}(a) shows that for $E_{So}=0.5$, the regular 
trajectory starting from $Q=1.0$ remains in the positive-$Q$ region because
a particle cannot go over the potential barrier of $\Delta=1.0$.
For $E_{So}=0.8$, chaotic trajectories may go to the negative-$Q$ region 
with a help of force from bath [Eq. \ref{eq:E4}]. Figure \ref{fig6}(d) shows that
when $E_{So}$ is too large compared to $\Delta$ ($E_{So}/\Delta=1.2$), 
the trajectory again becomes regular, going between positive- and negative-$Q$ regions.

\begin{figure}
\begin{center}
\includegraphics[keepaspectratio=true,width=100mm]{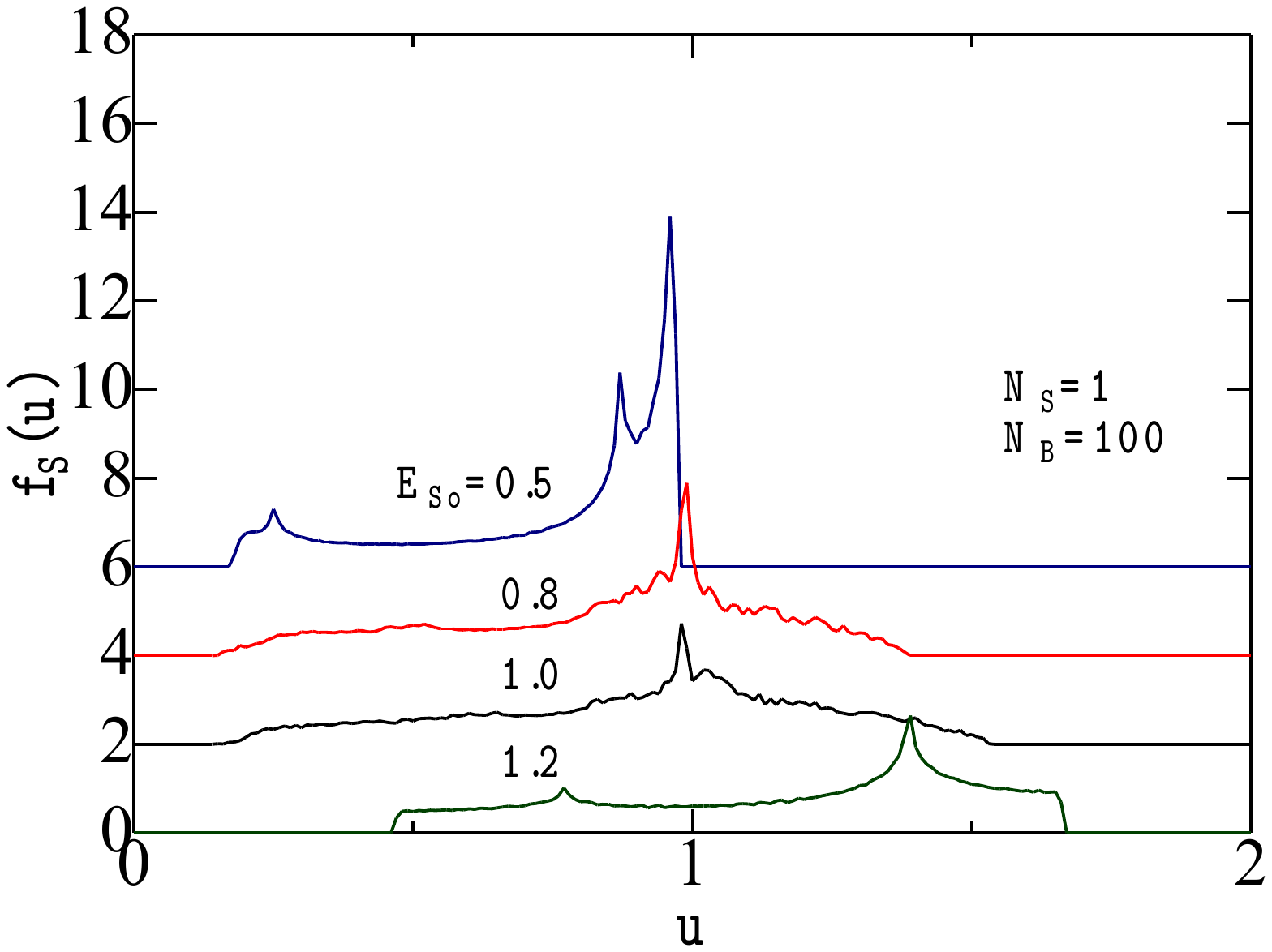}
\end{center}
\caption{
(Color online) 
System energy distributions $f_S(u)$ for $E_{So}=0.5$, 0.8, 1.0 and $1.2$ with $N_S=1$, 
$N_B=100$, $T=1.0$ and $c_o=1.0$, curves being successively shifted upward by two 
for clarity of figures. 
}
\label{fig7}
\end{figure}

Figure \ref{fig7} shows the system energy distribution $f_S(u)$
for various $E_{So}$.
$f_S(u)$ moves upward as $E_{So}$ is increased.
It is noted that peak positions of $f_S(u)$ for $E_{So}=0.5-1.0$
locate at $u \simeq 1.0$ while that for $E_{So}=1.2$ locates at $u \simeq 1.35$.

\subsection{Stationary energy probability distributions}

\begin{figure}
\begin{center}
\includegraphics[keepaspectratio=true,width=120mm]{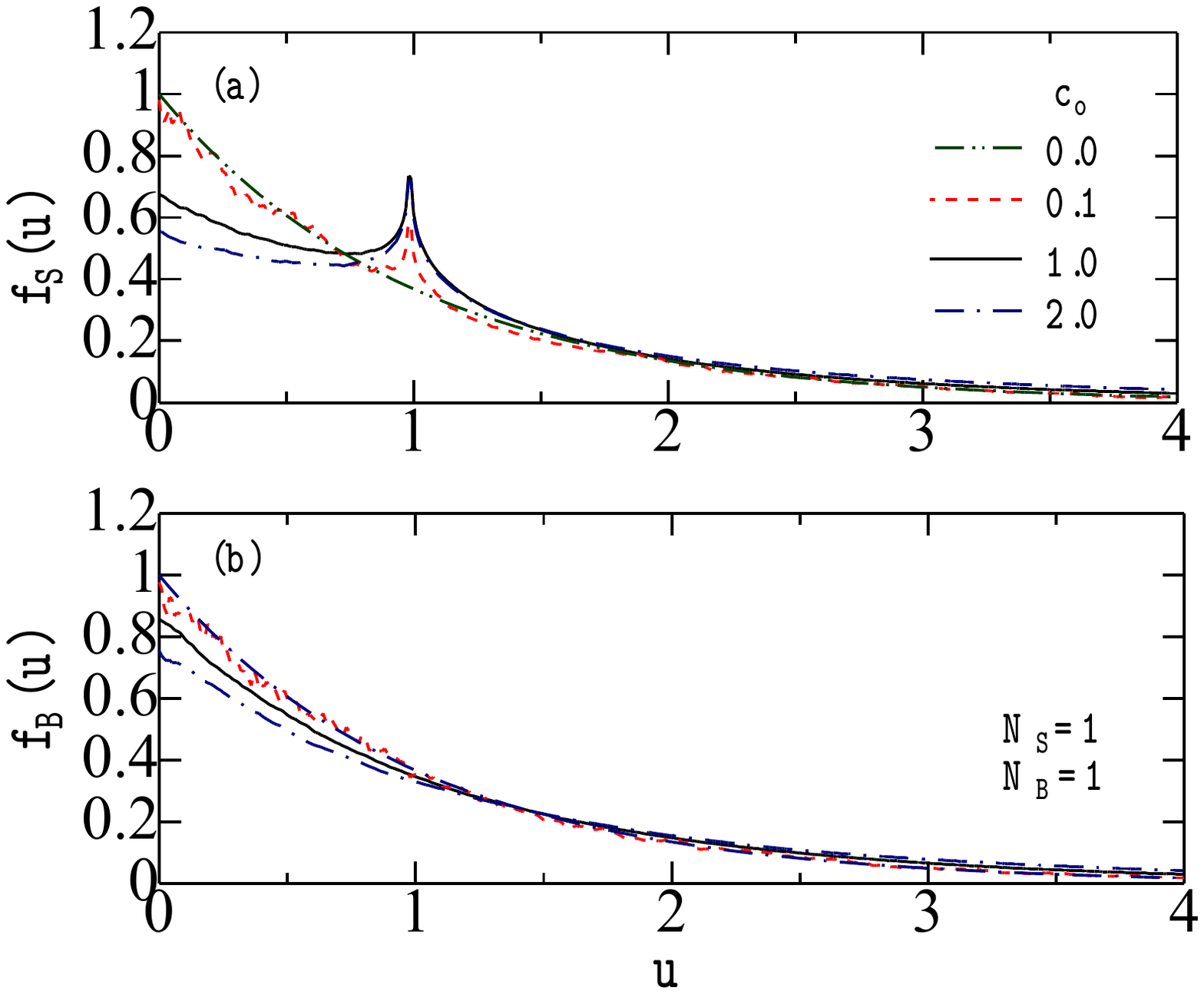}
\end{center}
\caption{
(Color online) 
Stationary distributions of (a) $f_S(u)$ and (b) $f_B(u)$ 
for various $c_o$: $c_o=0.0$ (double-chain curves), 0.1 (dashed curves),
1.0 (solid curves) and 2.0 (chain curves)
obtained by 10 000 runs  with $N_S=N_B=1$ and $T=1.0$. 
}
\label{fig8}
\end{figure}

In this subsection, we will study stationary energy probability distributions of system 
and bath which are averaged over $N_r$ (=10 000) runs stating 
from different initial conditions. 
Assuming that the system and bath are in the equilibrium states 
with $T=u_B=u_S$ at $t < 0.0$, we first generate exponential derivatives of 
initial system energies $\{ E_j \}$:
$p(E_j) \propto \exp(- \beta E_j)$ ($j=1$ to $N_S N_r$) for our DSs
where $\beta=1/k_B T$. 
A pair of initial values of $Q_j(0)$ and $P_j(0) $ 
for a given $E_j$ is randomly chosen such that they meet the condition 
given by $E_j=P_j(0)^2/2 M+V(Q_j(0))$. 
The procedure for choosing initial values of $q_n(0)$ and $p_n(0)$ 
is the same as that adopted in the preceding subsection \cite{Hasegawa11}. 
We have discarded results for $t < 200$ in our DSs performed for $t=0$ to 1000.

Before discussing cases where $N_S$ and $N_B$ may be greater than unity,
we first study a pedagogical simple case of $N_S=N_B=1$: a particle 
with double-well potential is subjected to a single harmonic oscillator. 
Double-chain curves in Fig. \ref{fig8}(a) and \ref{fig8}(b) show energy distributions 
of the system [$f_S(u)$] and bath [$f_B(u)$], respectively, with $c_o=0.0$, 
where $u=u_S$ ($u=u_B$) for the system (bath).
Both $f_S(u_S)$ and $f_B(u_B)$ follow the exponential distribution because 
the assumed initial equilibrium states of decoupled system and bath persist at $t \ge 0.0$.
When they are coupled by a weak coupling of $c_o=0.1$ at $t \geq 0.0$, 
$f_S(u)$ and $f_B(u)$ almost remain exponential distributions except for 
that $f_S(u)$ has a small peak at $u=1.0$, as shown by dashed curve in Fig. \ref{fig8}(a).  
This peak has been realized in Figs. \ref{fig2}(d) and \ref{fig3}(d).
It is due to the presence of a potential barrier with $\Delta=1.0$ 
in double-well potential because the peak at $u=1.0$ in $f_S(u)$ is realized 
even when $T \neq 1.0$, as will be discussed later in {\it 4. Effect of $T$} (Fig. \ref{fig12}). 
This peak is developed for stronger couplings of $c_o=1.0$ and 2.0, 
for which magnitudes of $f_S(u)$ at small $u$ are decreased, as shown by solid and chain curves
in Figs. \ref{fig8}(a) and \ref{fig8}(b).

\subsubsection{Effect of $c_o$}

\begin{figure}[htbp]
\begin{center}
\includegraphics[keepaspectratio=true,width=120mm]{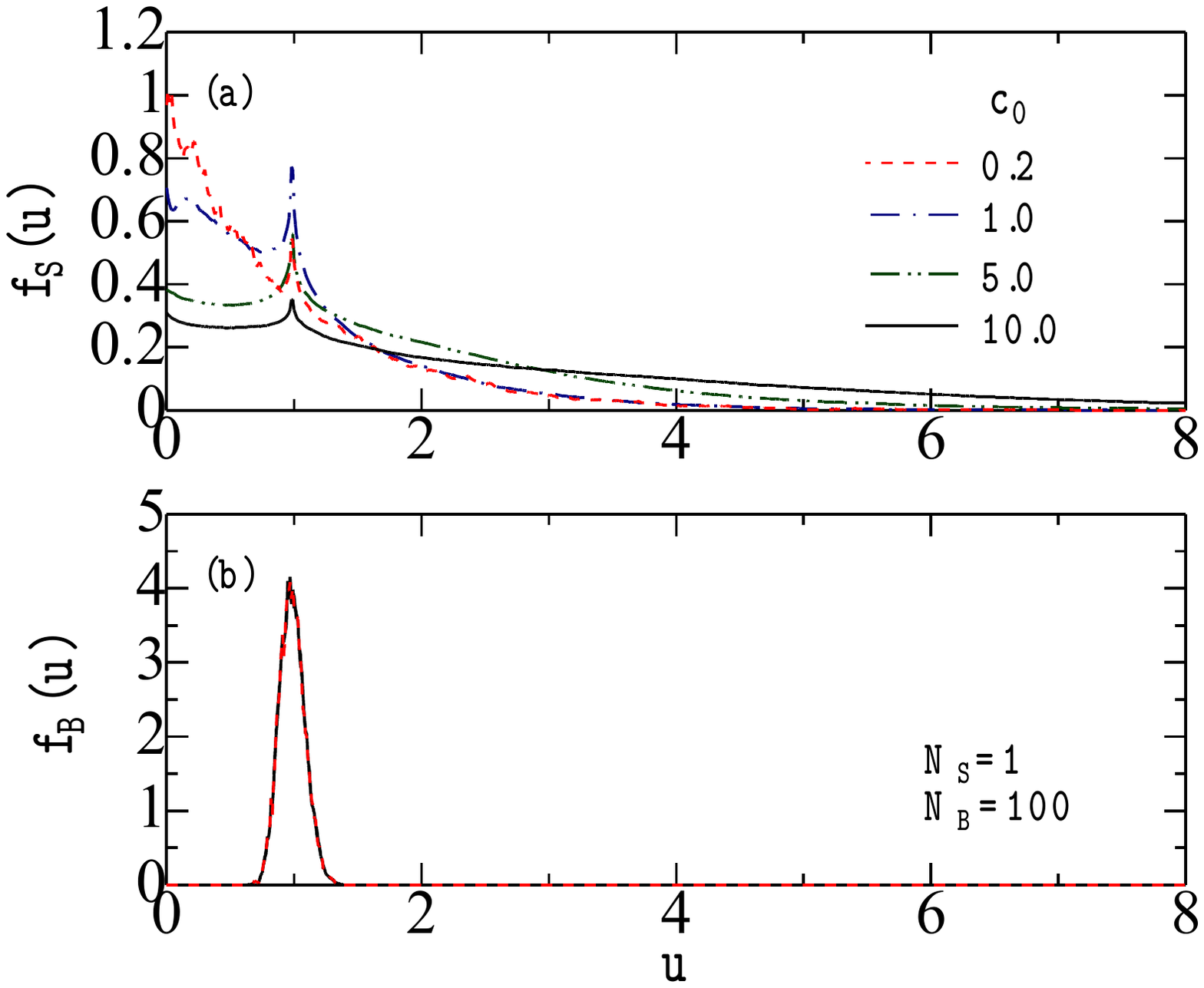}
\end{center}
\caption{
(Color online) 
Stationary distributions of (a) $f_S(u)$ and (b) $f_B(u)$ 
for various $c_o$: $c_o=0.2$ (dashed curves), 1.0 (chain curves), 5.0 (double-chain curves) 
and 10.0 (solid curves) with $N_S=1$, $N_B=100$ and $T=1.0$. 
}
\label{fig9}
\end{figure}

We change the coupling strength of $c_o$.
Figures \ref{fig9}(a) and \ref{fig9}(b) show $f_S(u)$ and $f_B(u)$, respectively,
for $c_o=0.2$, 1.0, 5.0 and 10.0 with $N_S=1$, $N_B=100$ and $T=1.0$.
$f_S(u)$ for $c_o=0.2$ nearly follows the exponential distribution.
When $c_o$ becomes larger, magnitudes of $f_S(u)$ at $u < 1.0$ are decreased 
while that at $u > 1.0$ is increased. In particular, the magnitude of
$f_S(0)$ is more decreased for larger $c_o$.

\subsubsection{Effect of $\omega_n$ distributions}

\begin{figure}
\begin{center}
\includegraphics[keepaspectratio=true,width=120mm]{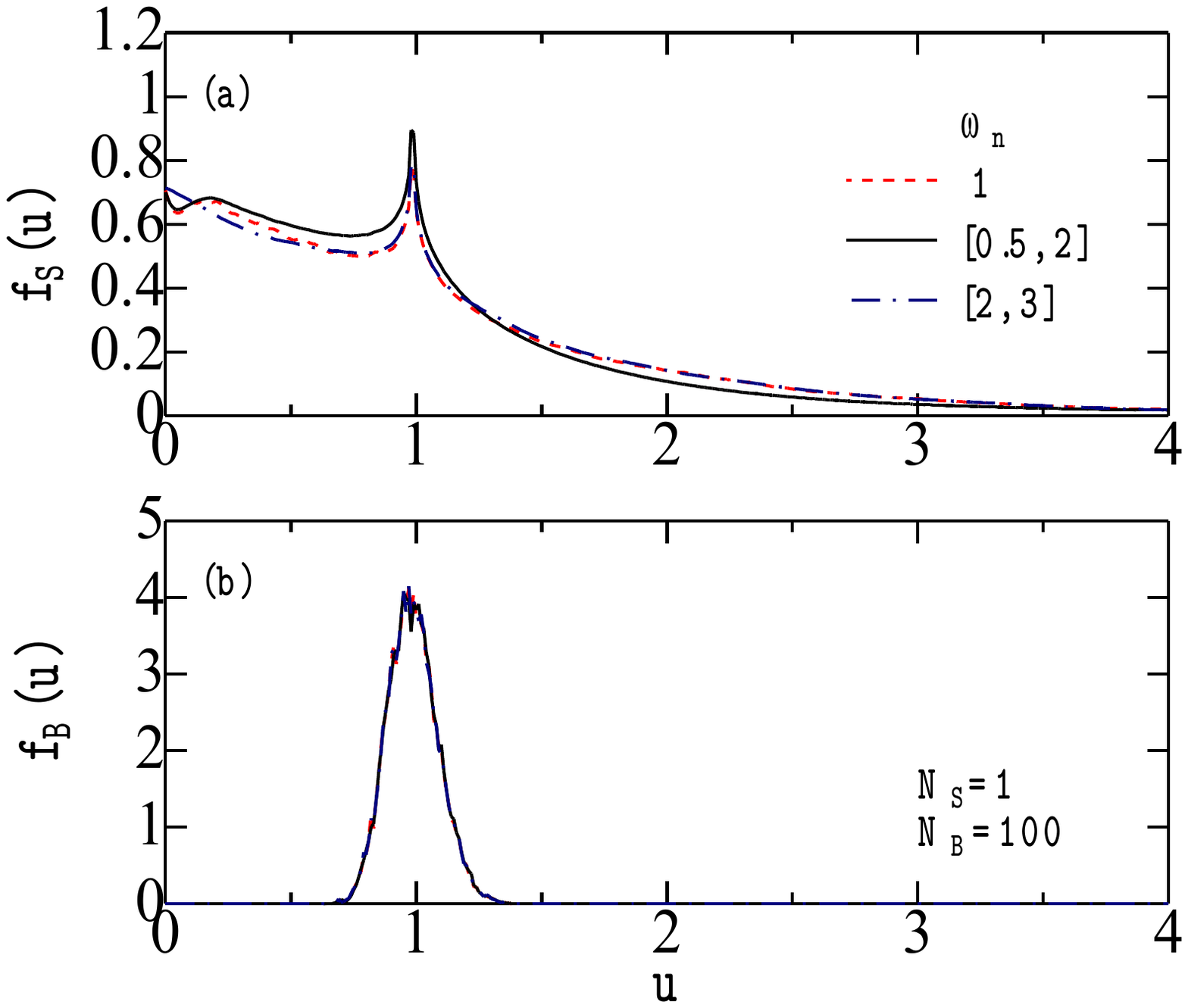}
\end{center}
\caption{
(Color online) Stationary distributions of (a) $f_S(u)$ and (b) $f_B(u)$ 
for various distributions of $\{ \omega_n\}$: 
$\omega=1.0$ (dashed curves), $\omega_n \in [0.5, 2.0]$ (solid curves)
and $\omega_n \in [2.0, 3.0]$ (chain curves) with $N_S=1$, $N_B=100$,
$T=1.0$ and $c_o=1.0$. 
}
\label{fig10}
\end{figure}

Although we have assumed $\omega_n=1.0$ in bath oscillators,
we will examine the effect of their distribution, taking into account
two kinds of random distributions given by
$\omega_n\in [0.5, 2.0]$ and $\omega_n \in [2.0, 3.0]$.
From calculated results shown in Figs. \ref{fig10}(a) and \ref{fig10}(b), we note 
that $f_S(u)$ and $f_B(u)$ are not much sensitive to the distribution
of $\{ \omega_n \}$ in accordance with our previous calculation
for harmonic oscillator system \cite{Hasegawa11,Note1}.
This conclusion, however, might not be applied to the case of infinite bath 
where distribution of $\{ \omega_n \}$ becomes continuous distribution.
Ref. \cite{Wei09} reported that the relative position 
between oscillating frequency ranges of system and bath is very important
for a thermalization of the harmonic oscillator system subjected to finite bath.

\subsubsection{Effect of $N_B$}

\begin{figure}
\begin{center}
\includegraphics[keepaspectratio=true,width=120mm]{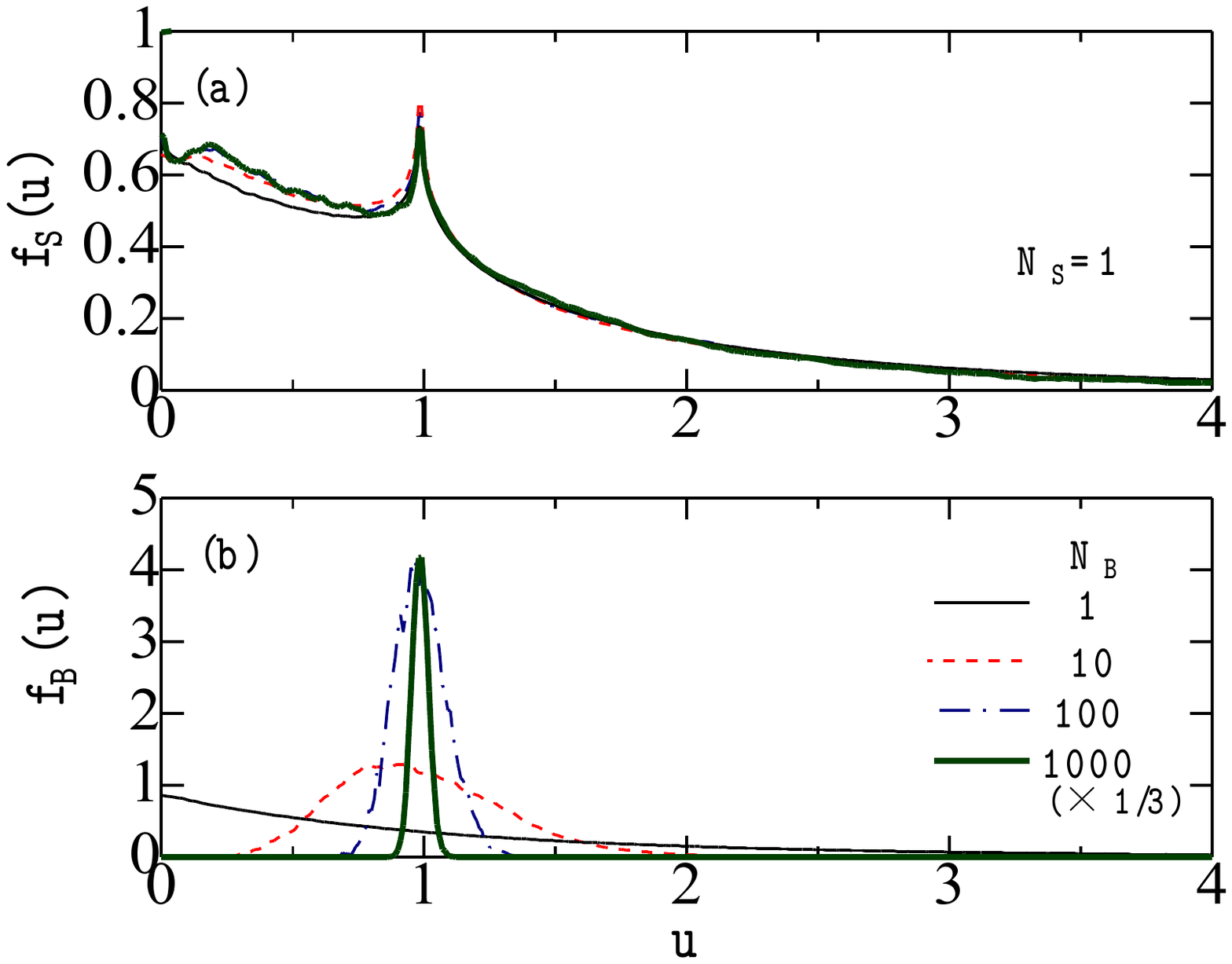}
\end{center}
\caption{
(Color online) Stationary distributions of (a) $f_S(u)$ and (b) $f_B(u)$ 
for various $N_B$: $N_B=1$ (solid curves), 10 (dashed curves),  
100 (chain curves) and 1000 (bold solid curves) with $N_S=1$, $T=1.0$ and $c_o=1.0$.
$f_B(u)$ for $N_B=1000$ is multiplied by a factor of $1/3$. 
}
\label{fig11}
\end{figure}

We have calculated $f_S(u)$ and $f_B(u)$, changing $N_B$ but with fixed
$N_S=1$, whose results are shown in Figs. \ref{fig11}(a) and \ref{fig11}(b). 
For larger $N_B$, the width of $f_B(u)$ becomes narrower as expected.
However, shapes of $f_S(u)$ are nearly unchanged for all cases of $N_B=1$, 10, 100 and 1000.

\subsubsection{Effect of $T$}

\begin{figure}
\begin{center}
\includegraphics[keepaspectratio=true,width=120mm]{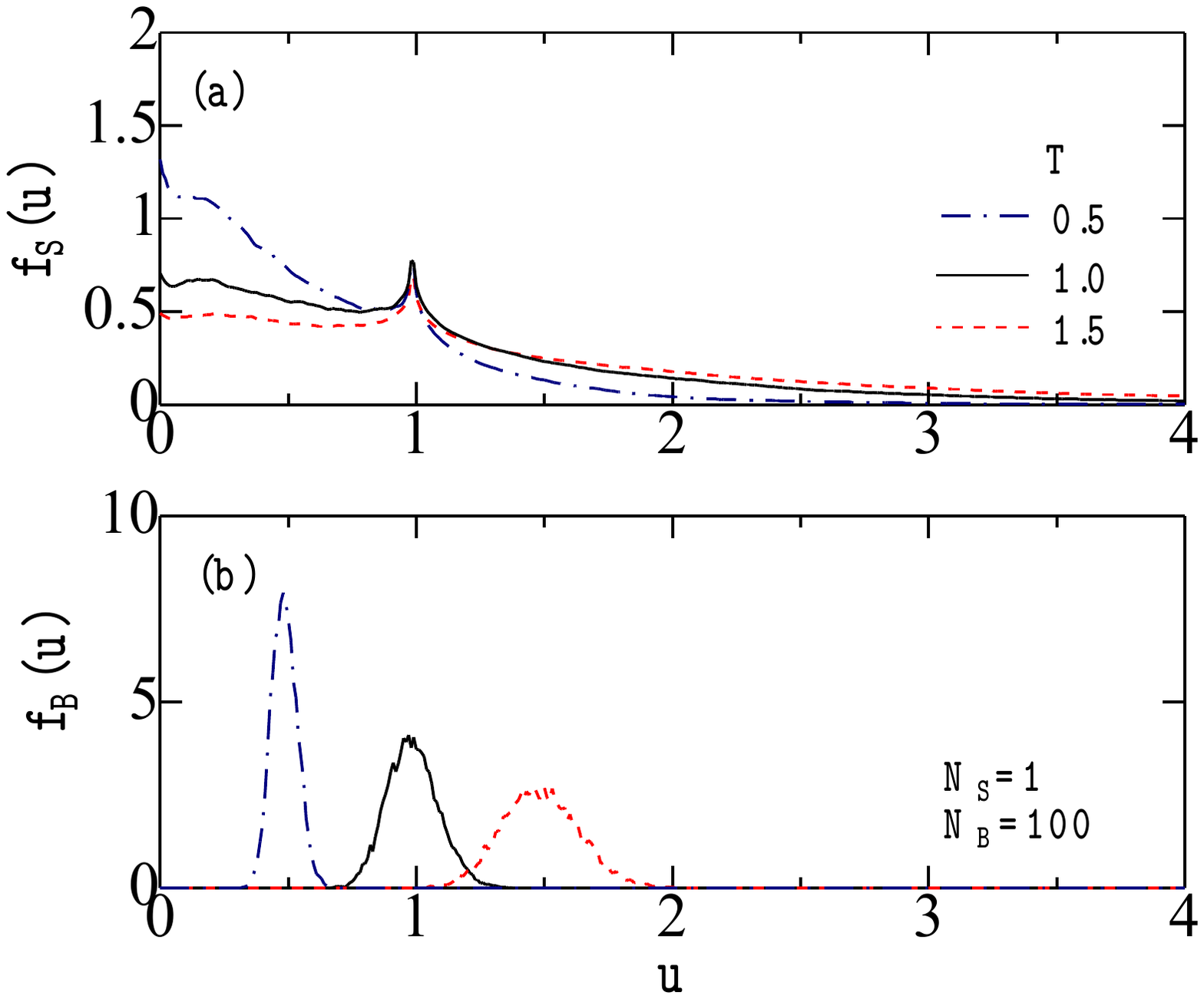}
\end{center}
\caption{
(Color online) Stationary distributions of (a) $f_S(u)$ and (b) $f_B(u)$ 
for various $T$: $T=0.5$ (chain curves), 
1.0 (solid curves) and 1.5 (dashed curves) with $N_S=1$, $N_B=100$ and $c_o=1.0$. 
}
\label{fig12}
\end{figure}

We change the temperature of the bath. 
Figures \ref{fig12}(a) and \ref{fig12}(b) show $f_S(u)$ and $f_B(u)$, respectively,
for $T=0.5$, 1.0 and 1.5 with $N_S=1$, $N_B=100$ and $c_o=1.0$.
When $T$ is decreased (increased), positions of $f_B(u)$ move to lower (higher)
energy such that mean values of $u_B$ correspond to $T$.
For a lower temperature of $T=0.5$, magnitude of $f_S(u)$ at $u < 1.0$ is increased
while that at $u > 1.0$ is decreased. 
The reverse is realized for higher temperature of $T=1.5$.
We should note that the peak position in $f_S(u)$ at $u=1.0$ is not
changed even if $T$ is changed because this peak is related to
the barrier with $\Delta=1.0$ of the double-well potential.

\subsubsection{Effect of $N_S$}

\begin{figure}
\begin{center}
\includegraphics[keepaspectratio=true,width=120mm]{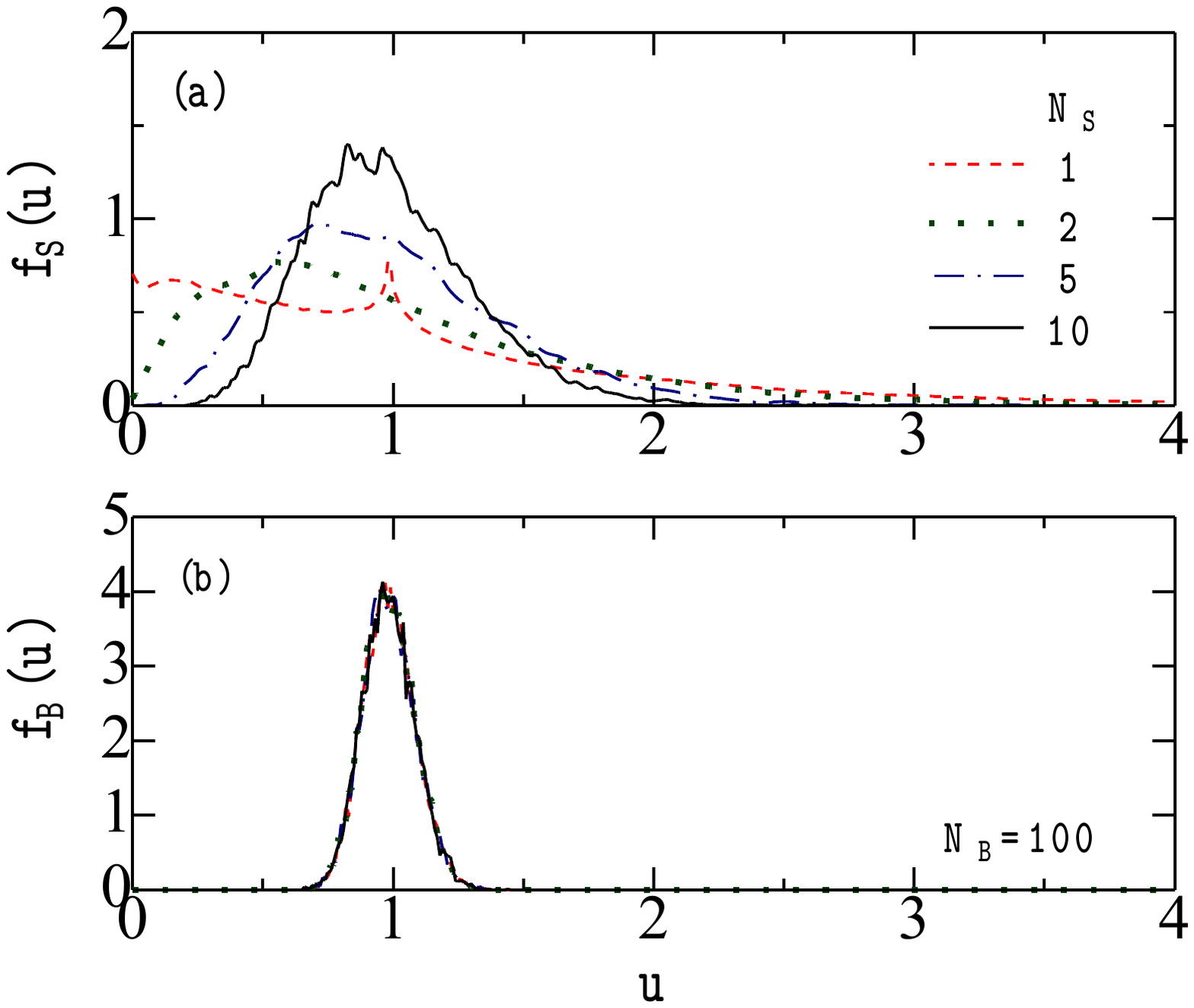}
\end{center}
\caption{
(Color online) Stationary distributions of (a) $f_S(u)$ and (b) $f_B(u)$ 
for various $N_S$: $N_S=1$ (dashed curves), 2 (doted curves),
5 (chain curves) and 10 (solid curves) with $N_B=100$, $T=1.0$ and $c_o=1.0$. 
}
\label{fig13}
\end{figure}

Although $N_S=1$ has been adopted so far, we will change $N_s$ 
to investigate its effects on stationary energy distributions.
Figure \ref{fig13}(a) shows $f_S(u)$ for $N_S=1$, 2, 5 and 10.
$f_S(u)$ for $N_S=1$ shows an exponential-like distribution
with $f_S(0) \neq 0$ at $u=0.0$.
In contrast, $f_S(u)$ vanishes at $u=0.0$ for $N_S=2$, 5 and 10.
Figure \ref{fig13}(a) shows that shapes of $f_S(u)$ much depend on $N_S$ 
while those of $f_B(u)$ are almost unchanged in Fig. \ref{fig13}(b).

\section{Discussion}
\subsection{Analysis of stationary energy distributions}
Our DSs in the preceding section have shown that $f_S(u)$ depends mainly on $N_S$, $c_o$ and $T$
while $f_B(u)$ depends mostly on $N_B$ and $T$ for $N_S \ll N_B$. 
We will try to analyze $f_S(u)$ and $f_B(u)$ in this subsection.
It is well known that when variables of $x_i$ ($i=1-N$) are independent and follow
the exponential distributions with the same mean, the distribution of
its sum: $X=\sum_i x_i$ is given by the $\Gamma$ distribution.
Then for an uncoupled system ($c_o=0.0$), $f_S(u)$ and $f_B(u)$ are expressed by 
the $\Gamma$ distribution $g(u)$ given by \cite{Hasegawa11}
\begin{eqnarray}
f_{\eta}(u) &=& \frac{1}{Z_{\eta}}\;u^{a_{\eta}-1}\:e^{-b_{\eta} \:u} 
\equiv g(u; a_{\eta}, b_{\eta}), 
\label{eq:B1} 
\end{eqnarray}
with
\begin{eqnarray}
a_{\eta} &=& N_{\eta}, \hspace{1cm} b_{\eta}= N_{\eta} \beta, 
\label{eq:B2}\\
Z_{\eta} &=& \frac{\Gamma(a_{\eta})}{b_{\eta}^{a_{\eta}} },
\label{eq:B3}
\end{eqnarray}
where $\eta=$S and B for a system and bath, respectively, and
$\Gamma(x)$ is the gamma function. 
In the limit of $N_S=1$, the $\Gamma$ distribution reduces to the exponential distribution.
Mean ($\mu_{\eta}$) and variance ($\sigma_{\eta}^2$) of the $\Gamma$ distribution
are given by
\begin{eqnarray}
\mu_{\eta} &=& \frac{a_{\eta}}{b_{\eta}}, 
\hspace{1cm} \sigma_{\eta}^2 =\frac{a_{\eta}}{b_{\eta}^2}, 
\label{eq:B4}
\end{eqnarray}
from which $a_{\eta}$ and $b_{\eta}$ are expressed 
in terms of $\mu_{\eta}$ and $\sigma_{\eta}$
\begin{eqnarray}
a_{\eta} &=& \frac{\mu_{\eta}^2}{\sigma_{\eta}^2}, 
\hspace{1cm} b_{\eta} =\frac{\mu_{\eta}}{\sigma_{\eta}^2}. 
\label{eq:B5}
\end{eqnarray}

\begin{figure}
\begin{center}
\includegraphics[keepaspectratio=true,width=100mm]{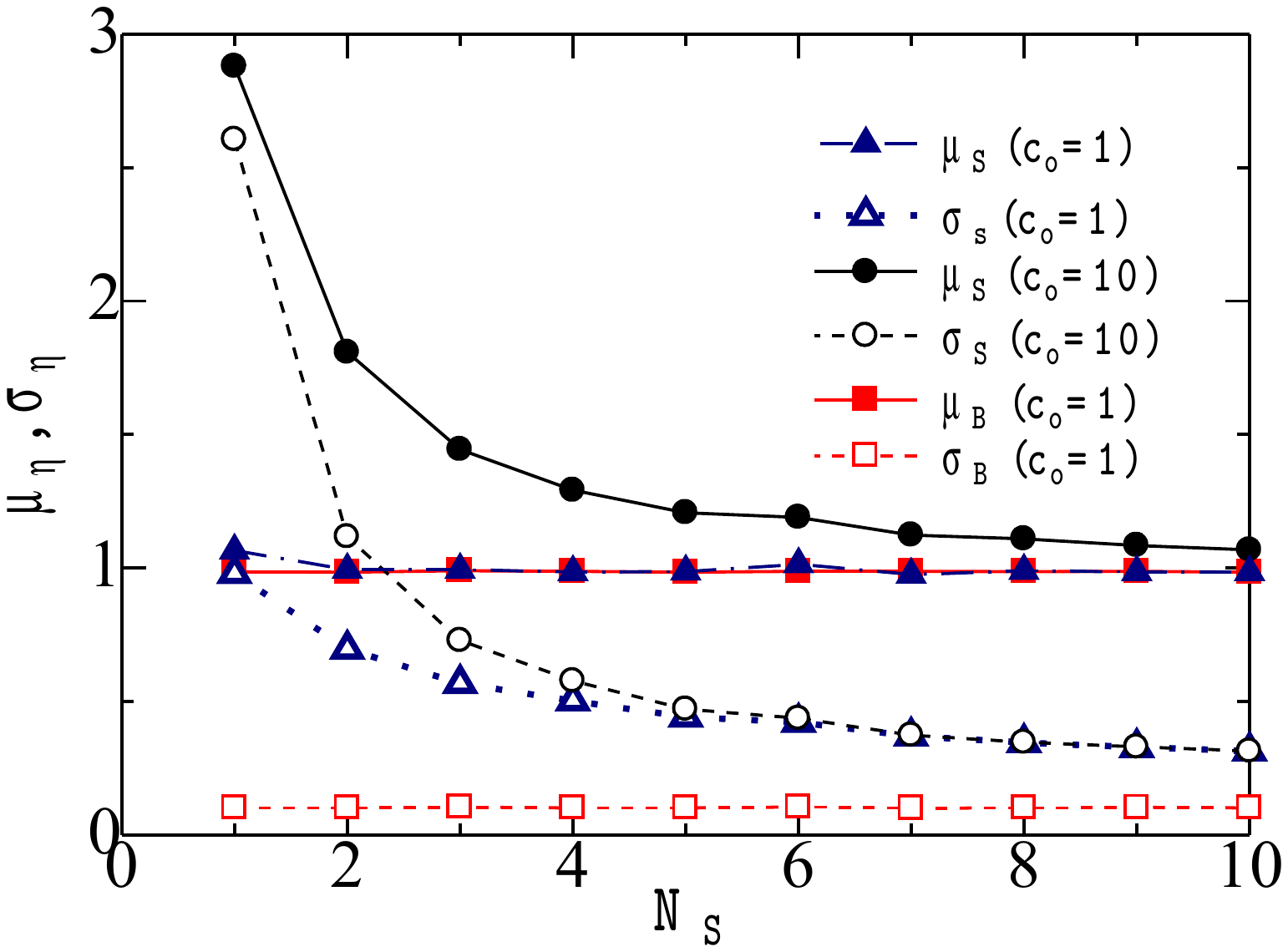}
\end{center}
\caption{
(Color online) $N_S$ dependences of $\mu_{\eta}$ and $\sigma_{\eta}$
of system ($\eta=S$) and bath $(\eta=B)$ with $N_B=100$ and $T=1.0$:
filled (open) triangles denote $\mu_S$ ($\sigma_S$) with $c_o=1.0$:
filled (open) circles express $\mu_S$ ($\sigma_S$) with $c_o=10.0$:
filled (open) squares show $\mu_B$ ($\sigma_B$) with $c_o=1.0$.
}
\label{fig14}
\end{figure}

\begin{figure}
\begin{center}
\includegraphics[keepaspectratio=true,width=120mm]{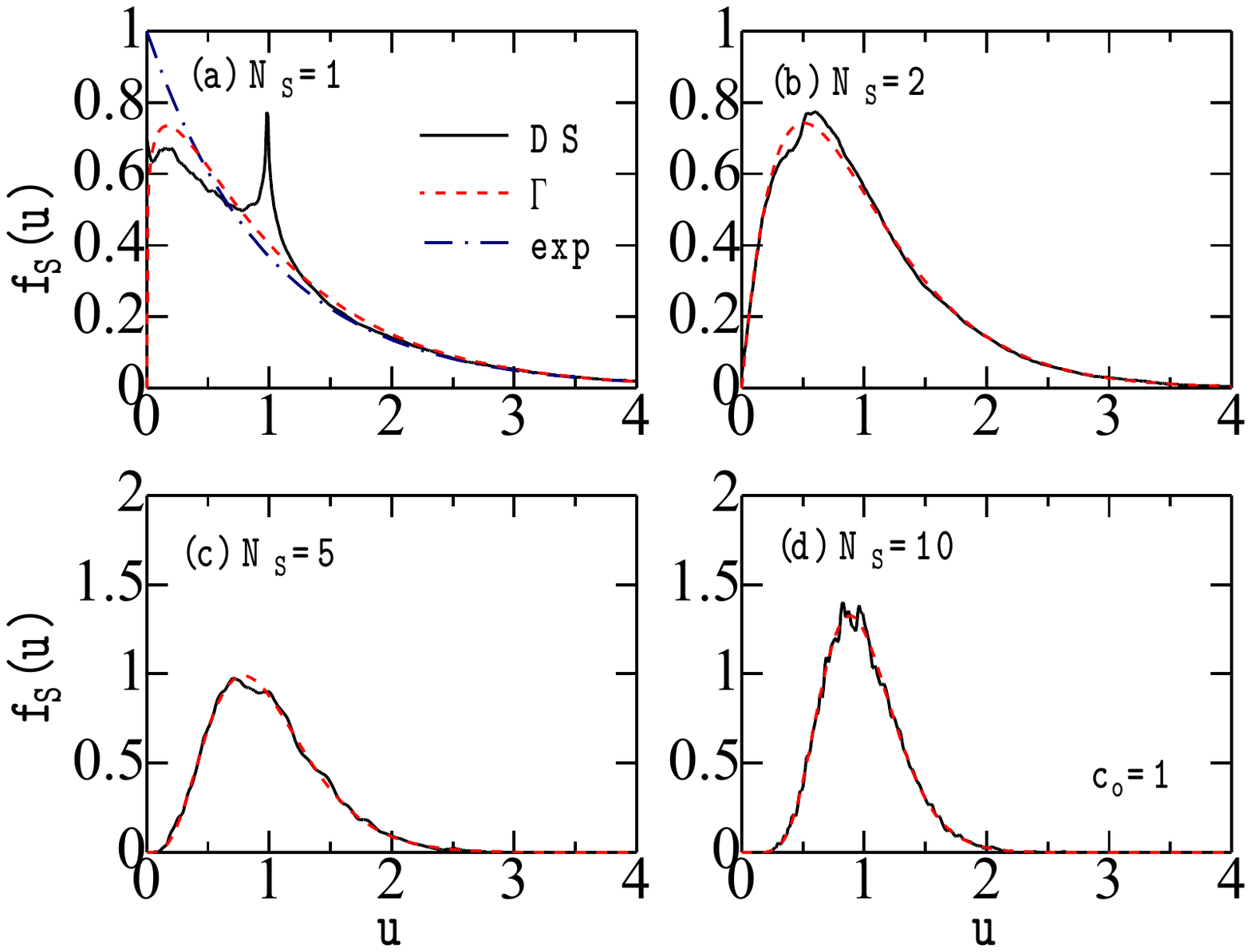}
\end{center}
\caption{
(Color online) $u$ dependences of $f_S(u)$ for (a) $N_S=1$, (b) $N_S=2$,
(c) $N_S=5$ and (d) $N_S=10$ with $T=1.0$, $c_o=1.0$ and $N_B=100$ obtained by 
DSs (solid curves): dashed and chain curves express $\Gamma$ and exponential distributions,
respectively (see text).
}
\label{fig15}
\end{figure}

\begin{figure}
\begin{center}
\includegraphics[keepaspectratio=true,width=120mm]{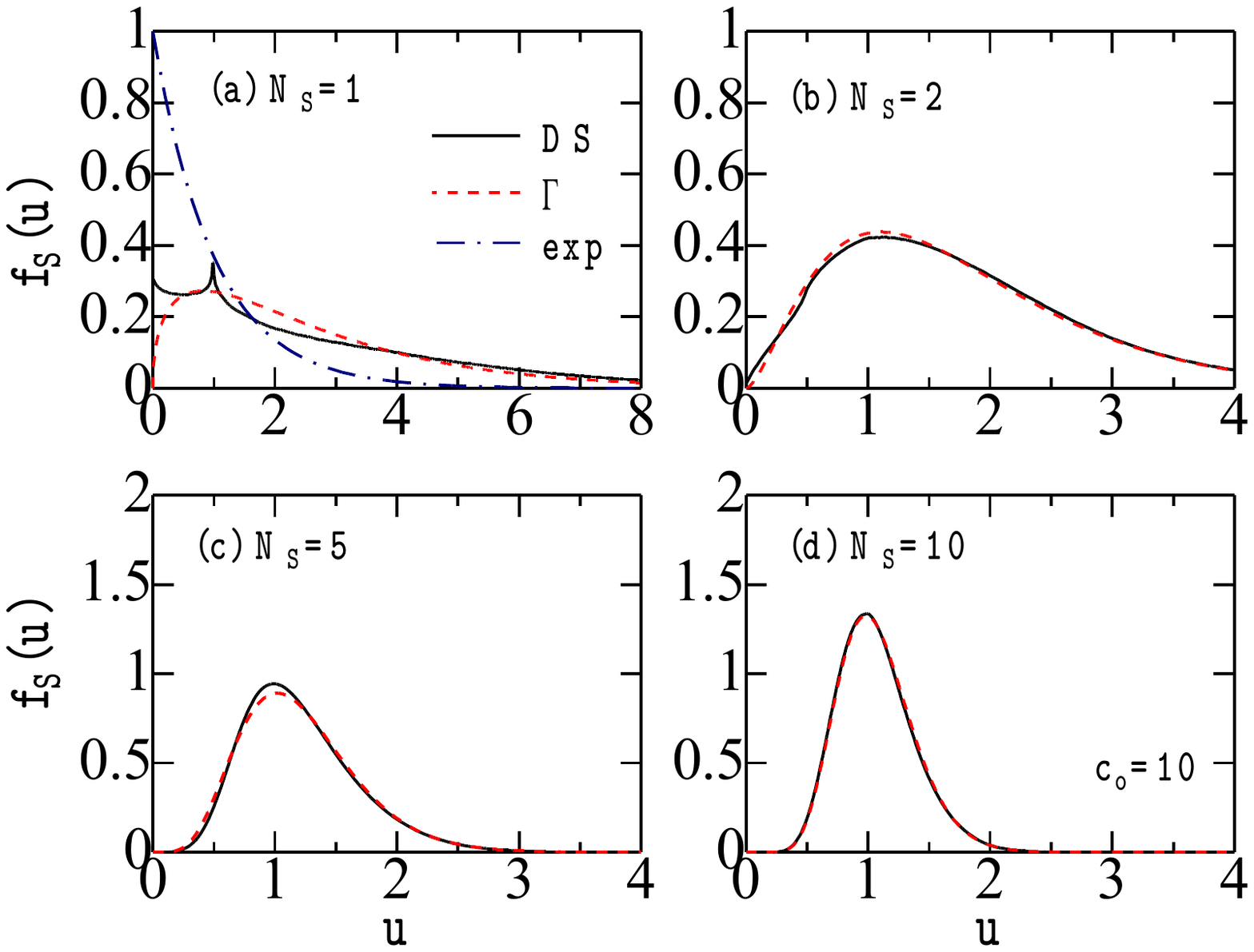}
\end{center}
\caption{
(Color online) $u$ dependences of $f_S(u)$ for (a) $N_S=1$, (b) $N_S=2$,
(c) $N_S=5$ and (d) $N_S=10$ with $T=1.0$, $c_o=10.0$ and $N_B=100$ obtained by DSs (solid curves):
dashed and chain curves express $\Gamma$ and exponential distributions,
respectively (see text).
}
\label{fig16}
\end{figure}

We have tried to evaluate $f_S(u)$ and $f_B(u)$ for the coupled system ($c_o \neq 0.0$) as follows: 
From mean ($\mu_{\eta}$) and root-mean-square (RMS) ($\sigma_{\eta}$) calculated by DSs,
$a_{\eta}$ and $b_{\eta}$ are determined by Eq. (\ref{eq:B5}), with which
we obtain the $\Gamma$ distributions for $f_S(u)$ and $f_B(u)$. 
Filled and open squares in Fig. \ref{fig14} show $\mu_B$ and $\sigma_B$,
respectively, as a function of $N_S$. 
We obain $\mu_B=1.0$ and $\sigma_B=0.1$ nearly independently of $N_S$,
which yield $a_B=b_B=100.0$ in agreement with Eq. (\ref{eq:B2}). 
Filled and open triangles in Fig. \ref{fig14} express the $N_S$ dependence of
$\mu_S$ and $\sigma_S$ obtained by DSs with $c_o=1.0$, $N_B=100$ and $T=1.0$. 
Calculated mean and RMS values of $(\mu_S, \sigma_S)$  
are $(1.07, 0.98)$, $(0.99, 0.70)$, $(0.99, 0.44)$
and $(0.99, 0.319)$ for $N_S=1$, 2, 5 and 10, respectively,
for which Eq. (\ref{eq:B5}) yields $(a_S, b_S)=(1.18, 1.11)$, $(2.04, 2.05)$, 
$(4.50, 5.06)$ and $(9.82, 9.97)$. 
These values of $a_S$ and $b_S$ are not so different from $N_S$ and $N_S \beta$
given by Eq. (\ref{eq:B2}).
We have employed the $\Gamma$ distribution with these parameters $a_S$ and $b_S$
for our analysis of $f_S(u)$ having been shown in Fig. \ref{fig13}(a).
Dashed curves in Figs. \ref{fig15}(a)-(d) express calculated $\Gamma$ distributions,
which are in fairly good agreement with $f_S(u)$ plotted by solid curves,
except for $N_S=1$ for which $g(0) = 0.0$ because $a_S=1.18 > 1.0$
while $f_S(0) \neq 0.0$.

Similar analysis has been made for another result obtained with a larger $c_o=10.0$
for $N_B=100$ and $T=1.0$. $N_S$-dependences of calculated $\mu_S$ and $\sigma_S$ 
are plotted by filled and open circles, respectively, in Fig. \ref{fig14}. 
Calculated $(\mu_S, \sigma_S)$ are $(2.88, 2.61)$, $(1.81, 1.12)$, 
$(1.21, 0.47)$ and $(1.07, 0.31)$ for $N_S=1$, 2, 5 and 10, respectively,
which lead to $(a_S, b_S)=(1.22, 0.42)$, $(2.62, 1.45)$, $(6.57, 5.44)$
and $(11.54, 10.81)$ by Eq. (\ref{eq:B5}).
Obtained $a_S$ and $b_S$ are rather different from
$N_S$ and $N_S \beta$ given by Eq. (\ref{eq:B2}).
Dashed curves in Figs. \ref{fig16}(a)-(d) show $\Gamma$ distributions
with these parameters, which may approximately explain $f_S(u)$ obtained by DSs 
in the {\it phenomenologically} sense, except for $N_S=1$ for which
$g(0)=0.0$ but $f_S(0) \neq 0.0$.

We note in Fig. \ref{fig15}(a) or \ref{fig16}(a) that an agreement between 
$g(u)$ and $f_S(u)$ with $N_S=1$ is not satisfactory.
We have tried to obtain a better fit between them, by using the $q$-$\Gamma$ distribution
$g_q(u)$ given by \cite{Hasegawa11}
\begin{eqnarray}
g_q(u) &=& \frac{1}{Z_q} \:u^{a-1} \:e_q^{-b u},
\end{eqnarray}
with
\begin{eqnarray}
e_q^x &=& [1+(1-q)x]_+^{1/(1-q)},
\end{eqnarray}
where $[y]_+= \max(y,0)$ and $Z_q$ is the normalization factor. Note that
$g_q(u)$ reduces to the $\Gamma$ distribution in the limit of $q \rightarrow 1.0$.
Although the $q$-$\Gamma$ distribution was useful for $f_S(u)$ of harmonic-oscillator systems 
subjected to finite bath \cite{Hasegawa11}, 
it does not work for $f_S(u)$ of double-well systems.
This difference may be understood from a comparison between $f_S(u)$ for $N_S=1$ of a double-well system
shown in Fig. \ref{fig15}(a) [or \ref{fig16}(a)] and its counterpart
of a harmonic oscillator system shown in Fig. 9(a) of Ref. \cite{Hasegawa11}.
Although the latter shows an exponential-like behavior with a monotonous decrease 
with increasing $u$, the former with a characteristic peak at $u=1.0$ 
cannot be expressed by either 
the exponential, $\Gamma$, or $q$-$\Gamma$ distribution. 

\subsection{Analysis of stationary position distributions}

\begin{figure}
\begin{center}
\includegraphics[keepaspectratio=true,width=100mm]{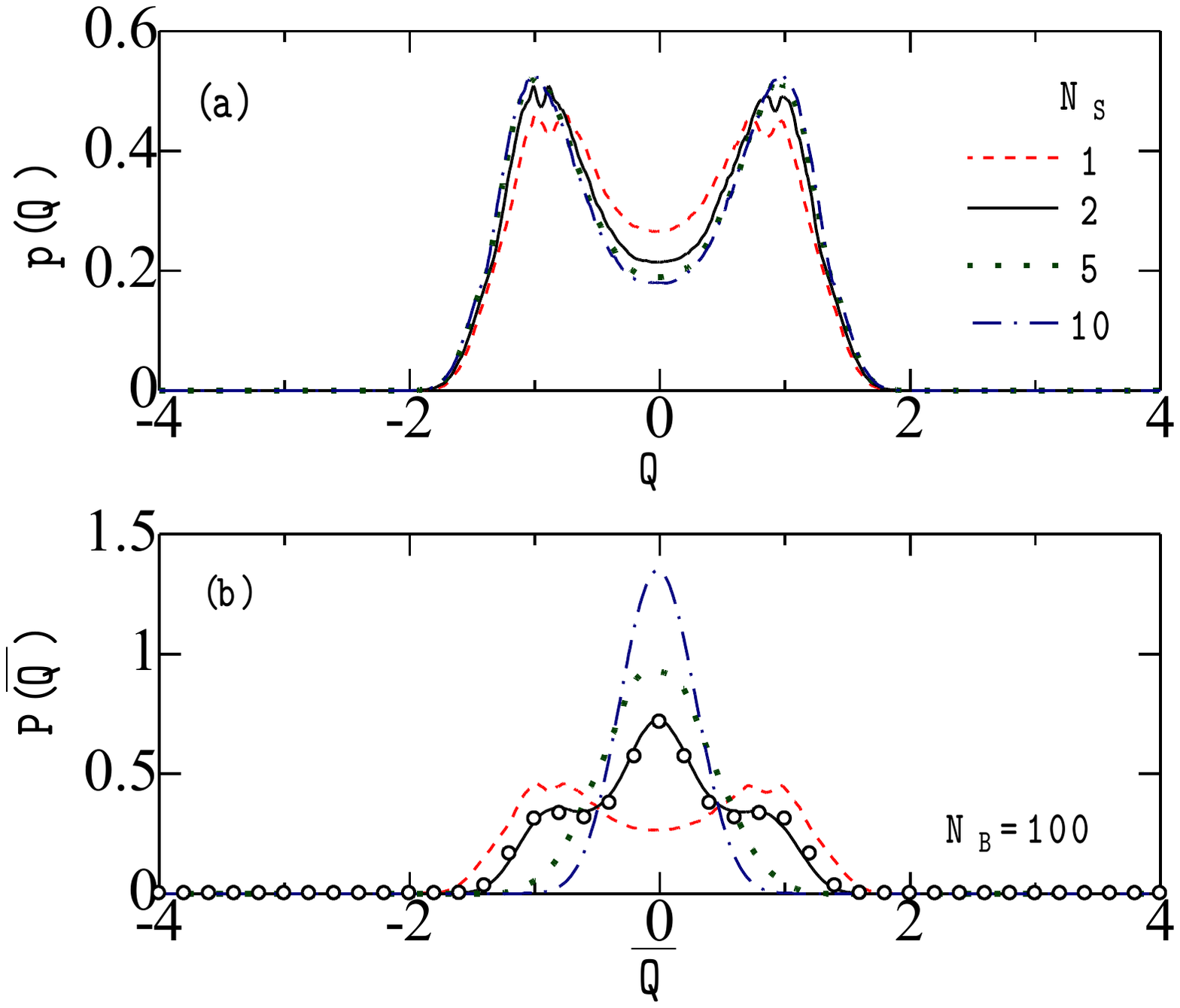}
\end{center}
\caption{
(Color online) Stationary distributions of (a) $p(Q)$ as a function of particle position
$Q$ and (b) $P(\bar{Q})$ as a function of the averaged position $\bar{Q}$ 
for various $N_S$: $N_S=1$ (dashed curve), 2 (solid curve), 5 (dotted curve) 
and 10 (chain curve) with $N_B=100$, $T=1.0$ and $c_o=1.0$. 
Open circles in (b) express an analytical result obtained by Eq. (\ref{eq:C2})
with $N_S=2$. 
}
\label{fig17}
\end{figure}

We have studied also the $N_S$ dependence of stationary position distributions of
$p(Q)$ and $P(\bar{Q})$, where $Q$ denotes the position of a particle
in the system and $\bar{Q}$ expresses the averaged position given by
\begin{eqnarray}
\bar{Q} &=& \frac{1}{N_S} \sum_{k=1}^{N_S} Q_k.
\label{eq:C1}
\end{eqnarray}
Figures \ref{fig17}(a) and \ref{fig17}(b) show $p(Q)$ and $P(\bar{Q})$, respectively, 
obtained by DSs for various $N_S$ with $N_B=100$, $T=1.0$ and $c_o=1.0$.
For $N_S=1$, we obtain $p(Q)=P(\bar{Q})$ with the characteristic double-peaked structure. 
We note, however, that $P(\bar{Q})$ is different from $p(Q)$ for $N_S > 1$
for which $P(\bar{Q})$ has a single-peaked structure despite the double-peaked $p(Q)$.
This is easily understood as follows: 
For example, in the case of $N_S=2$, two particles in the system
mainly locate at $Q_k=1.0$ or $Q_k=-1.0$ ($k=1, 2$)
which yields the double-peaked distribution of $f_S(Q)$.
However, the averaged position of $\bar{Q}=(Q_1+Q_2)/2$ will be dominantly $\bar{Q}=0.0$,
which leads to a single-peaked $P(\bar{Q})$. The situation is the same also for $N_S > 2$.  

Theoretically $P(\bar{Q})$ may be expressed by
\begin{eqnarray}
P(\bar{Q}) &=& \int \cdot\cdot \int \prod_{k=1}^{N_S}\:dQ_k  
\;\exp\left[-\beta V(Q_k) \right]
\:\delta\left(\bar{Q}-N_S^{-1} \sum_{k=1}^{N_S} Q_k \right).
\label{eq:C2}
\end{eqnarray}
$P(\bar{Q})$ numerically evaluated for $N_S=2$ is plotted 
by open circles in Fig. \ref{fig17} which are in good agreement with the solid curve
expressing $P(\bar{Q})$ obtained by DS.
It is impossibly difficult to numerically evaluate Eq. (\ref{eq:C2}) for $N_S \geq 3$.
In the limit of $N_S \rightarrow \infty$, $P(\bar{Q})$ reduces to the Gaussian distribution
according to the central-limit theorem.
This trend is realized already in the case of $N_S=10$ in Fig. \ref{fig17}(b).

\section{Concluding remarks}

We have studied the properties of classical double-well systems coupled to
finite bath, employing the ($N_S+N_B$) model \cite{Hasegawa11}
in which $N_S$-body system is coupled to $N_B$-body bath.
Results obtained by DSs have shown the following:

\noindent
(i) Chaotic oscillations are induced in the double-well system
coupled to finite bath in the absence of external forces for
appropriate model parameters of $c_o$, $N_B$, $T$, $\{ \omega_n \}$ and $E_{So}$,

\noindent
(ii) Among model parameters, 
$f_S(u)$ depends mainly on $N_S$, $c_o$ and $T$
while $f_B(u)$ depends on $N_B$ and $T$ for $N_S \ll N_B$,
 
\noindent
(iii) $f_S(u)$ for $N_S > 1$ obtained by DSs may be phenomenological expressed 
by the $\Gamma$ distribution, 

\noindent
(iv) $f_S(u)$ for $N_S=1$ with $c_o \neq 0.0$ cannot be described 
by either the exponential, $\Gamma$, or $q$-$\Gamma$ distribution, 
although that with $c_o=0.0$ follows the exponential distribution, and

\noindent
(v) The dissipation is not realized in the system energy for DSs at $t=0-1000$
with $N_S=1-100$ and $N_B=10-1000$.

\noindent
The item (i) is in consistent with chaos in a closed classical
double-well system driven by external forces \cite{Reichl84},
although chaos is induced without external forces
in our open classical double-well system. 
This is somewhat reminiscent of chaos induced by quantum noise 
in the absence of external force in closed quantum double-well systems \cite{Pattanayak94}.
Effects of induced chaos in the item (i) are not apparent in $f_S(u)$ 
because $u$ ($=u_S$) is ensemble averaged over 10 000 runs (realizations)
with exponentially distributed initial system energies.
Items (ii) and (v) are the same as in the harmonic-oscillator system coupled to
finite bath \cite{Hasegawa11}.
The item (v) suggests that for the energy dissipation of system,
we might need to adopt a much larger $N_B$ ($\gg 1000$) \cite{Poincare}. 
The item (iv) is in contrast to $f_S(u)$ for $N_S=1$ 
in the open harmonic-oscillator system which may be approximately accounted for
by the $q$-$\Gamma$ distribution \cite{Hasegawa11}. 
It would be necessary and interesting to make a quantum extension of our study
which is left as our future subject.

\begin{acknowledgments}
This work is partly supported by
a Grant-in-Aid for Scientific Research from 
Ministry of Education, Culture, Sports, Science and Technology of Japan.  
\end{acknowledgments}


\end{document}